\documentclass[12pt,a4paper,twoside]{article}

\usepackage{graphicx,hyperref,bm,url,microtype,color,bbm}
\usepackage{amsfonts,amssymb,amsthm,amsmath,enumitem}
\usepackage[english]{babel}
\usepackage{natbib}
\usepackage{comment} 
\usepackage{tikz}
\usepackage{float} 

\usepackage{multirow}

\usepackage{caption}
\usepackage{subcaption}

\usepackage[a4paper,text={16cm,23.5cm},centering]{geometry}
\usepackage[compact,small]{titlesec}
\usepackage[utf8]{inputenc}

\def \d{\text{\rm d}}
\def \Emp{\mathbb{E}}
\def \E{\text{\rm E}}
\def \P{{\text{\rm P}}}

\def \B{\mathbb{B}}
\newcommand{\sgn}{\mbox{\rm sgn}}
\newcommand{\R}{\ensuremath{\mathbb{R}}}

\newcommand{\cp}{\ensuremath{ \xrightarrow[]{\P}  }}
\newcommand{\cw}{\ensuremath{ \rightarrow_{\text{\rm w}}}}

\newcommand{\G}{\mathbf{G}}

\DeclareMathOperator*{\boldtheta}{{\boldsymbol{\theta}}}

\DeclareMathOperator*{\Cov}{Cov}

\newcommand{\bpsi}{\boldsymbol\psi}

\newcommand{\btheta}{\boldsymbol\theta}
\newcommand{\bcero}{\mathbf 0}

\newcommand{\bV}{\mathbf V}

\newcommand{\bbF}{\mathbb F}
\newcommand{\bbG}{\mathbb G}
\newcommand{\bbR}{\mathbb R}
\newcommand{\bbX}{\mathbb X}

\newcommand{\bbZ}{\mathbb Z}

\definecolor{violet}{rgb}{0.7,0,0.6}
\theoremstyle{definition}

\theoremstyle{theorem}

\newtheorem{theorem}{Theorem}
\newtheorem{proposition}{Proposition}
\newtheorem{corollary}{Corollary}

\newtheorem{theoremAppendix}{Theorem A.\ignorespaces}

\newtheorem{lemmaAppendix}{Lemma A.\ignorespaces}

\newenvironment{keywords}
    {\vspace*{3mm}
    {\noindent{}\textit{Keywords\/:}}
        \nopagebreak\small}
        {}

\title{\textbf{\Large Bootstrap tests for almost goodness-of-fit
}}
\author{Amparo Ba\'illo$^1$ and Javier C\'{a}rcamo$^{2,}$\footnote{Corresponding author: Javier C\'{a}rcamo \newline  \quad Affiliation: Departamento de Matem\'{a}ticas, Universidad del Pa\'{\i}s Vasco, Aptdo. 644, 48080 Bilbao (SPAIN)\newline E-mail Address: javier.carcamo@ehu.eus} \\
{\small{$^{1}$ Departamento de Matem\'{a}ticas, Universidad Aut\'{o}noma de Madrid, 28049 Madrid (SPAIN)}}\\
{\small{$^{2}$ Departamento de Matem\'{a}ticas, Universidad del Pa\'{\i}s Vasco, Aptdo. 644, 48080 Bilbao (SPAIN)}}
}
\date{\today}

\begin{document}

\maketitle

\begin{abstract}
\noindent
We introduce the \textit{almost goodness-of-fit} test, a procedure to assess whether a (parametric) model provides a good representation of the probability distribution generating the observed sample.
Specifically, given a distribution function $F$ and a parametric family $\mathcal{G}=\{ G(\boldsymbol{\theta}) : \boldsymbol{\theta} \in \Theta\}$, we consider the testing problem
\[
H_0: \| F - G(\boldsymbol{\theta}_F) \|_p \geq \epsilon \quad \text{vs} \quad H_1: \| F - G(\boldsymbol{\theta}_F) \|_p < \epsilon,
\]
where $\epsilon>0$ is a margin of error and $G(\boldsymbol{\theta}_F)$ denotes a representative of $F$ within the parametric class. The approximate model is determined via an M-estimator of the parameters. 
The methodology also quantifies the percentage improvement of the proposed model relative to a non-informative (constant) benchmark. The test statistic is the $\mathrm{L}^p$-distance between the empirical distribution function and that of the estimated model. We present two consistent, easy-to-implement, and flexible bootstrap schemes to carry out the test. The performance of the proposal is illustrated through simulation studies and analysis and real-data applications.
\end{abstract}
\begin{keywords}
Bootstrap consistency; Empirical processes; Equivalence test; Mixtures; Model validation; Relevant hypotheses.
\end{keywords}




\section{Introduction and motivation} \label{Section.Introduction}


Goodness-of-fit (GoF) tests are classical problems in statistical inference.
They are used to decide whether the true distribution underlying a sample follows a specific model or, more generally, belongs to a parametric family of distributions. The final goal of a GoF test is to check whether the model is a reasonably good approximation to the unknown distribution of the population.
However, practically every GoF test places this statement in the null hypothesis \(H_0\), thus being able to establish statistical evidence only for the lack of fit of the population to the model and not for the actual goodness of fit.

In the statistical literature, there has been interest in tests whose alternative hypothesis is the GoF of the model to the true distribution (see, e.g., \cite{Wellek21}).
In this case, to obtain a well-posed test, the class of distributions included in \(H_1\) has to be a suitable enlargement of the model that would be used in the null hypothesis of traditional GoF tests. This means that the new alternative hypothesis states that the true unknown distribution is within a specified positive ``margin'' of the potential model.
For instance, in biostatistics, the term \textit{equivalence test} encompasses statistical tests whose alternative hypothesis is that the generating distribution of the sample does not deviate from the proposed model by more than an equivalence margin (\cite{Romano05}, \cite{Wellek10}).

There are different approaches to overcome the limitations of the traditional GoF procedures. \cite{Davies-2014} introduced the concept of \textit{approximate model}: a model that generates samples resembling the observed data. The notion of approximation is related to a measure of closeness, and this often requires a metric. For the two-sample problem, \cite{Munk-Czado-1998} considered tests for a trimmed version of the Mallows distance between two cumulative distribution functions (cdf) to assess the similarity between them. In the context of multinomial GoF tests, \cite{Liu-Lindsay-2009}
define a tubular neighborhood given by multinomial distributions whose Kullback-Leibler distance to the proposed model does not exceed a pre-specified tolerance level.
\cite{AE-B-CA-M-2012} and \cite{del Barrio-2020} exploit probability trimmings and contamination neighborhoods to assess similarity between distributions within the framework of robustness.

A related line of research focuses on testing \textit{relevant} or \textit{precise} hypotheses, where the null hypothesis involves a nonzero lower bound on the discrepancy between distributions or parameters. In this setting, the goal is not to test for exact equality, but rather to determine whether the discrepancy exceeds a given threshold. This idea appears prominently in \cite{Berger-Delampady}, who analyze the philosophical and statistical implications of testing sharp hypotheses within a Bayesian framework. In the same spirit, \cite{Baringhaus-Henze} consider the Cram\'{e}r--von Mises distance to define a \textit{neighbourhood-of-$F_0$ validation test}. \cite{Dette-Sen} also propose consistent testing procedures for relevant hypotheses.

In this paper we consider what we call \textit{almost goodness-of-fit} (AGoF in short) tests, a general framework to validate (i.e., find evidence) that the data are well described by the selected model and to quantify the goodness of the approximation. The alternative hypothesis contemplates that the distribution of the variable of interest might not be exactly equal to the proposed model, but is ``very close'' to it in terms of an appropriate distance. The degree of dissimilarity allowed is quantified by a parameter $\epsilon > 0$, the \textit{margin}, which can be set in advance. Specifically, for a fixed $\epsilon > 0$ and a probability metric (or semi-metric) $d$,
we are interested in tests of the form:
\begin{equation}\label{Tests.AGoF}
\text{(a)}\quad
\begin{cases}
H_0: & d(F,F_0)\ge \epsilon,\\
H_1: &  d(F,F_0)< \epsilon,
\end{cases}
\quad \text{(b)}\quad
\begin{cases}
H_0: & d(F,G(\boldsymbol{\theta}_F))\ge \epsilon,\\
H_1: &  d(F,G(\boldsymbol{\theta}_F))< \epsilon,
\end{cases}
\end{equation}
where $F$ is the unknown cdf of the observed variable \(X\), $F_0$ is a known/specified cdf and
\begin{equation}\label{parametric-family}
\mathcal{G}=\{ G(\boldsymbol{\theta})\equiv G(x;\boldsymbol{\theta}) : x\in\R\text{ and }\boldsymbol{\theta} \in \Theta\subset \R^k\}, \quad k \in\mathbb{N},
\end{equation}
is a family of cdf depending on a $k$-dimensional parameter. Here, $\boldsymbol{\theta}_F \in \Theta$ is determined by solving an
M-estimation problem (see Section~\ref{Subsection.Model}).

The rejection of the null hypothesis in \eqref{Tests.AGoF}(a) means that there is statistical evidence that the population distribution is \textit{almost} $F_0$, while, in the usual GoF tests with simple null hypothesis, non-rejection of \(H_0:F=F_0\) only means that there is not enough evidence against the equality of the distributions.
\cite{Wellek21} considered an alternative hypothesis as in \eqref{Tests.AGoF}(a) composed of Lehmann alternatives of \(F_0\) whose supremum distance from \(F_0\) is bounded by a specified small margin.
The alternative hypothesis in \eqref{Tests.AGoF}(b) states that the distribution of \(X\) is at most within a margin \(\epsilon\) of the parametric family \(\mathcal{G}\). In practice, this is likely the most useful test of the two in \eqref{Tests.AGoF}. For this reason, we focus on problem \eqref{Tests.AGoF}(b) since \eqref{Tests.AGoF}(a) can be treated similarly.

In this work we propose a bootstrap rejection region for the test \eqref{Tests.AGoF}(b) 
when \(d\) is the \(\mathrm{L}^p\)-metric between the cdf's and under very general assumptions (satisfied by many usual parametric models).
The methodology also allows for the swap of hypotheses in \eqref{Tests.AGoF}. In this way, we can deal with the problem $H_0:  d(F,G(\boldsymbol{\theta}_F))\le \epsilon$ versus $H_1:  d(F,G(\boldsymbol{\theta}_F)) >  \epsilon$, which can be viewed as a relaxed version of the classic GoF test. As we discuss below, in this context the choice of the distance $d$ as the $\mathrm{L}^p$-metric has several advantages over the more usual supremum norm. 
In Section~\ref{Section.TestAGoF} we state the hypothesis test, propose a rejection region for it, and suggest how to use the asymptotic distribution of the test statistic to approximate this region in practice. We also give some indications on the choice and interpretation of the margin of error $\epsilon$ and introduce a quantity that measures the quality of the AGoF. In Section~\ref{Section.AsympDistr} we derive the limit distribution of the test statistic and prove the consistency of bootstrap approximations to the critical value of the rejection region. Section~\ref{Section.Simulations} illustrates the finite-sample performance of the AGoF testing procedure with a Monte Carlo study. In Section~\ref{Section.Data} we apply the AGoF test to two real data sets.
The proofs of the theoretical contributions are collected in the Appendix.

\section{Almost Goodness of Fit Tests} \label{Section.TestAGoF}

In this section we state the hypothesis test which is our main target, as well as the intuition behind the proposed rejection region.

\subsection{The choice of the proximity measure} \label{Subsection.Proximity}

As mentioned above, the concept ``almost'' in AGoF is necessarily accompanied by the idea of proximity. This translates into the use of a certain probability metric to quantify the differences between the observed data and the model under consideration. In statistics, the Kolmogorov or uniform distance is commonly used. Kolmogorov-Smirnov-type statistics are easy to understand and implement.   Moreover, in the simplest cases of GoF problems, this metric generates distribution-free methods.


In the particular case of AGoF tests, it is necessary to estimate the distance between the distribution of the population, which generally does not follow the model, and a representative within the model. For the uniform distance, the limiting distributions associated with the testing problems in \eqref{Tests.AGoF} cannot be treated as easily as in the case of the usual GoF; they are usually unwieldy, complex and non-Gaussian (see \cite{Raghavachari}). In addition, alternative computational procedures for dealing with difficult-to-treat distributions such as the bootstrap are not usually consistent when estimating the sup-norm, as it follows from \cite{Fang-Santos} and \cite{Carcamo2020}.


We propose to use $\mathrm{L}^p$-metrics (with $1 \le p < \infty$) to quantify the difference between the observed empirical distribution and the model. This choice has several advantages. First, the limit distribution of the associated statistic is, in general, more tractable. In fact, under mild assumptions, the asymptotic distribution is Gaussian. Moreover, as shown in \cite{Carcamo}, the $\mathrm{L}^p$-norms are Hadamard differentiable for $1 < p < \infty$; see \cite{Carcamo2020} for a precise definition. This property is crucial for applying the functional delta method, which is a key ingredient in the derivation of the asymptotic distribution of the test statistic (see Theorem~\ref{Theorem-probability-metric-parameters}). The Hadamard differentiability of the $\mathrm{L}^p$-norms for $1 < p < \infty$ also implies, under certain conditions, the consistency of standard bootstrap estimators of these distances (see \cite{Fang-Santos}). For the case $p = 1$, the associated norm is not fully Hadamard differentiable in general. However, a consistent and easy-to-compute bootstrap estimator is provided in \cite[Thm.~3]{Baillo-2024}.
In addition, we consider the $\mathrm{L}^p$-distances with respect to the Lebesgue measure rather than weighted versions such as in the Cram\'{e}r--von Mises setting as in \cite{Baringhaus-Henze}. The Lebesgue measure yields a more transparent interpretation of the discrepancy, avoiding the overemphasis on high-density regions of the model distribution. This is particularly relevant when the tails of the distribution are of interest, as model-based weighting schemes tend to downweight discrepancies in those regions.

The $\mathrm{L}^p$-metric also provides a way to control the relative importance assigned to the tails of the distributions in the approximate validation of the model: the larger the value of $p$, the less influence the tails will have. More generally, the $\mathrm{L}^p$ norms form a flexible family of metrics indexed by $p$, allowing the practitioner to tune the sensitivity of the test to different types of discrepancies: smaller values of $p$ emphasize global deviations, whereas larger values give more weight to pronounced local differences. The price to pay for using these norms (instead of other more commonly employed procedures) is that the underlying variables must satisfy additional integrability conditions to ensure that the corresponding statistics are well defined and converge. For example, the Cramér-von Mises test can be computed without imposing any integrability conditions on the variables involved, which may offer an advantage in certain settings.

Among the different proximity measures available in the literature, we choose to work with the $\mathrm{L}^p$-distance between the empirical distribution and the fitted model, measured with respect to the Lebesgue measure. This choice is motivated by several factors. First, while the Cram\'{e}r--von Mises distance is a classical option in goodness-of-fit testing, it only leads to a distribution-free test statistic in the simple univariate setting without parameter estimation. In the parametric or approximate GoF setting, the limiting distribution of such statistics depends intricately on both the underlying distribution and the parameter estimation procedure (see, e.g., \cite{Baringhaus-Henze}). In this context, the $\mathrm{L}^p$-distance does not entail a greater technical burden and provides additional interpretability. Indeed, integrating with respect to the Lebesgue measure ensures that the distance treats all regions of the domain uniformly, without biasing toward areas where the model density is higher. This is especially relevant in applications where discrepancies in the tails are important, as tail regions typically receive negligible weight under model-based measures such as $\d G(\boldsymbol{\theta})$. Finally, the $\mathrm{L}^p$ family offers a flexible class of metrics whose sensitivity can be adjusted through the parameter $p$, allowing users to emphasize different types of deviations between model and data.

\subsection{The model framework} \label{Subsection.Model}


The family \(\mathcal G\) in \eqref{parametric-family} is the potential model that might approximate the cdf \(F\) of the r.v. \(X\). To estimate the parameter $\btheta\in\Theta$, we assume that there exists a function \(\bpsi_{\btheta}\) such that
\[
\E_{G(\btheta)} \bpsi_{\btheta}(X) =  \int \bpsi_{\btheta} (x) \, \d G(x;\btheta) = \bcero, \quad \mbox{for all \(\btheta\in\Theta\)},
\]
where `$\E_{G(\btheta)}$' denotes the expectation with respect to the probability measure with cdf $G(\btheta)$. We do not impose that \(F\in\mathcal G\), but we assume that we can estimate the parameter \(\btheta\) with a sample drawn from \(F\). For this purpose, we use M-estimators (see \cite{Stefanski-Boos}) and work under the assumption that there exists \(\btheta_F\) in the interior of \(\Theta\) uniquely determined by the equations
\begin{equation} \label{theta_F}
    \E_{F} \bpsi_{\btheta_F}(X) = \int \bpsi_{\btheta_F} (x)\, \d F(x) = \bcero.
\end{equation}
For instance, if
we estimate $\boldtheta$ via maximum likelihood, then $\boldsymbol{\psi}_{\boldtheta}(x)=\partial \log g(x;\boldsymbol{\theta})/\partial \boldsymbol{\theta}$,
where $g(x;\boldsymbol{\theta})$ is the density function of $G(\boldsymbol{\theta})$. In such a case, $G(\boldtheta_F)$ is the projection of $F$ onto the family $\mathcal{G}$  using the Kullback-Leibler divergence; see the notion of \textit{misspecified model} in \cite[Example 5.25]{van der Vaart}.

Beyond maximum likelihood, many standard parameter estimators fall within the
M-estimation framework. For instance, the sample mean and variance are M-estimators
for location-scale families, including normal distributions.
In one-parameter exponential families, such as the exponential or Poisson distributions,
the natural estimating equations yield M-estimators that coincide with the classical
method-of-moments estimators based on sample means.
Similarly, for two-parameter families such as the gamma distribution,
estimating equations for the shape and scale parameters also define M-estimators.
These examples illustrate that the approach in \eqref{theta_F} covers a wide range of
classical parametric estimation problems.

\subsection{The AGoF test} \label{Subsection.AGoFtest}


When the metric \(d\) in \eqref{Tests.AGoF} is the $\mathrm{L}^p$-distance between \(F\) and its best representative in \(\mathcal G\), $G(\boldtheta_F)$, we get the AGoF test
\begin{equation}\label{AGoFspecific}
\begin{cases}
H_0: &  \|F-G(\btheta_F)\|_p \geq \epsilon, \\
H_1: &  \|F-G(\btheta_F)\|_p < \epsilon.
\end{cases}
\end{equation}
Here, \( \|f\|_p = (\int |f|^p)^{1/p} \) denotes the \(\mathrm{L}^p\)-norm of a function \(f\in \mathrm{L}^p=\mathrm{L}^p(\bbR)\).

\cite{Baringhaus-Henze} study the test \eqref{AGoFspecific} using the Cram\'{e}r--von Mises
distance in place of the usual $\mathrm{L}^p$-norm and focusing on the exponential family.
In contrast, our approach, based on M-estimators, accommodates a broad class of parametric
models simultaneously. 
The alternative hypothesis in \eqref{AGoFspecific} intuitively means that the distribution of \(X\) is well described by the model in \(\mathcal G\) up to an error, quantified by \(\|F-G(\btheta_F)\|_p\), of magnitude at most \(\epsilon\). If \(\epsilon\) is ``small enough'', then we might opt for \(\mathcal G\) as a satisfactory approximation to \(F\).
We discuss below how to choose and interpret the margin or error \(\epsilon\). 

In the following, we derive a rejection region for \eqref{AGoFspecific}. Let \(X_1,\ldots,X_n\) be a sample from \(X\) and let \(\bbF_n\) be the associated empirical distribution function, i.e.,
$$
\bbF_n(t)=\frac{1}{n}\sum_{i=1}^n 1_{\{X_i\le t\}},\quad n\in \mathbb{N},\quad t\in \R,
$$
where $1_A$ stands for the indicator function of the set $A$.
The M-estimator of \(\btheta_F\) is the solution \(\hat\btheta_n\) of the equations
\begin{equation} \label{Psi}
{\boldsymbol{\Psi}}_n(\boldtheta) = \E_{\bbF_n} \bpsi_{\btheta}(X)
= \frac{1}{n} \sum_{i=1}^n \bpsi_{\btheta}(X_i) = \bcero.
\end{equation}
For a significance level \(\alpha\), we propose the rejection region
\begin{equation*}
R = \{\|\bbF_n-G(\hat\btheta_n)\|_p<\epsilon-c(\alpha)\},
\end{equation*}
where \(c(\alpha)\) is chosen so that, asymptotically, the size of the test is bounded by \(\alpha\).
The test statistic is therefore $\|\bbF_n-G(\hat\btheta_n)\|_p$, and its normalized version is
\begin{equation}\label{normalized-statistic}
T_{n}(F,G({\boldtheta}_F),p) = \sqrt{n}(\|\bbF_n-G(\hat\btheta_n)\|_p-\|F-G(\btheta_F)\|_p).
\end{equation}

In Theorem~\ref{Theorem-probability-metric-parameters} of Section~\ref{Section.AsympDistr}, we derive the asymptotic distribution of the normalized statistic in \eqref{normalized-statistic}. Specifically, we show that
\begin{equation} \label{WeakLimit}
T_{n}(F,G({\boldtheta}_F),p) \cw T(F,G(\btheta_F),p),
\end{equation}
where `\(\cw\)' stands for weak convergence and the precise expression of the limit $T(F, G(\btheta_F), p)$ is given in Theorem~\ref{Theorem main}, through equations~\eqref{AsympDistr_p_equal_1} and~\eqref{AsympDistr_p_larger_1}. From this result, we obtain that
\[ \P\left(  \|\bbF_n-G(\hat\btheta_n)\|_p \le   \|F-G(\btheta_F)\|_p + Q_T(\alpha)/\sqrt{n} \right) \to \alpha,\quad \text{as } n\to\infty, \]
where $Q_T(\alpha) \equiv Q_{T(F, G(\btheta_F), p)}(\alpha)$ denotes the $\alpha$-quantile of the limit distribution in~\eqref{WeakLimit}. Therefore, the rejection region can be approximated by
\begin{equation} \label{c_alpha}
R_n = \{\|\bbF_n-G(\hat\btheta_n)\|_p<\epsilon-c_n(\alpha)\},\quad \text{with}\quad  c_n(\alpha)=-Q_T(\alpha)/\sqrt{n}.
\end{equation}
Observe that, for a given $\epsilon > 0$, the probability of rejecting $H_0$ in \eqref{AGoFspecific} is
\begin{equation}\label{probability.rejection}
    \P(\text{Reject } H_0) = \P\left(  T_n <  \sqrt{n}(\epsilon -  \|F-G(\btheta_F)\|_p) + Q_T(\alpha)  \right).
\end{equation}
 Therefore, from \eqref{probability.rejection}, we can derive the properties of the test associated with the rejection region \eqref{c_alpha}, which are summarized in the following proposition.

\begin{proposition}\label{Proposition.Properties} Let $\epsilon>0$ be fixed. For the testing problem \eqref{AGoFspecific}, the rejection region in \eqref{c_alpha} fulfills the following properties:
\begin{enumerate}[label=(\roman*), topsep=0mm, itemsep=1mm, parsep=0mm, align=left, labelwidth=*, leftmargin=-0.5 mm]
    \item Under $H_0$, if $\|F-G(\btheta_F)\|_p=\epsilon$, then $\P(\textrm{Reject } H_0) \to \alpha$, as $n\to\infty$.
    \item Under $H_0$, if $\|F-G(\btheta_F)\|_p>\epsilon$, then $\P(\textrm{Reject } H_0) \to 0$, as $n\to\infty$.
    \item Under $H_1$ ($\|F-G(\btheta_F)\|_p<\epsilon$), $\P(\textrm{Reject } H_0) \to 1$, as $n\to\infty$.
    \end{enumerate}
\end{proposition}

As a by-product of the convergence in \eqref{WeakLimit}, we can also obtain a symmetric  rejection region
\(\tilde R_n = \{\|\bbF_n-G(\hat\btheta_n)\|_p>\epsilon+c_n(1-\alpha)\}\),
for the dual AGoF test
\begin{equation*} 
\begin{cases}
H_0: & \|F-G(\btheta_F)\|_p \leq \epsilon \\
H_1: & \|F-G(\btheta_F)\|_p > \epsilon,
\end{cases}
\end{equation*}
where the null and alternative hypotheses have been interchanged with respect to \eqref{AGoFspecific}. Although we do not explore this alternative testing problem in detail, an analogue of Proposition~\ref{Proposition.Properties} can be derived for the dual AGoF test.

In Section~\ref{Section.AsympDistr} (Theorem~\ref{Theorem-probability-metric-parameters}), we derive the expression for the limit in \eqref{WeakLimit} and establish conditions (Corollary~\ref{Corollary normalidad}) under which it is Gaussian. We note that the quantity \( c_n(\alpha) \) in \eqref{c_alpha} depends on the underlying distribution \( F \) and the model \( G(\boldsymbol{\theta}_F) \), which are unknown in practice. For this reason, we prove in Corollary~\ref{Corollary.Bootstrap.Consistency.Lp} that, under suitable assumptions, it can be consistently approximated via bootstrap. This enables the implementation of the testing procedure whenever the M-estimator of the parameter can be computed and the family \(\mathcal{G}\) satisfies the conditions specified in Section~\ref{Section.AsympDistr}.

\subsection{The margin of error and a measure of AGoF} \label{Subsection.margin}

Regarding the natural question of how to choose the margin \(\epsilon\) in \eqref{AGoFspecific}, \cite{Wellek21} considers that it has to be discussed for each individual dataset and depends on the interests of the researcher dealing with the data. \cite{Liu-Lindsay-2009} give a detailed revision of this matter, but still consider it a delicate and complicated matter.

One possibility to avoid choosing a specific value for the margin of error is to determine the infimum of the \(\epsilon\) for which the null hypothesis in the AGoF test \eqref{AGoFspecific} is rejected at a given significance level $\alpha$. In other words, we propose to compute the \textit{minimum distance} (from $F$ to the model $\mathcal{G}$) at level $\alpha$ given by
\begin{equation}\label{minimum distance}
    \epsilon^*(\alpha)=\inf\{ \epsilon>0 : H_0 \text{  in \eqref{AGoFspecific} is rejected at level $\alpha$}\}.
\end{equation}
This quantity provides a measure of how ``good'' the model is when compared to other models (see \cite{del Barrio-2020} and the references therein).

Another relevant issue is to interpret the value $\epsilon^*(\alpha)$ in \eqref{minimum distance}. We aim at introducing an informative quantity relative to the \textit{quality} of the AGoF in terms of the $\mathrm{L}^p$-distance. Just as the value of the supremum distance has a clearer interpretation, it is not so simple to make a decision on the suitability of a model in terms of the $\mathrm{L}^p$-norms. In addition, it is convenient to have a normalized value (with values in $[0,1]$, for example) to measure the AGoF and compare  different models easily. 
We propose here an approach similar to the one to evaluate models using ANOVA. We consider the worst-case scenario to approximate $F$ as a model given by a constant variable equal to the mean  of $F$, say $\mu$. In a way, the distribution of a degenerate random variable with probability measure $\delta_\mu$ taking the value of the mean $\mu$ almost surely is the coarsest model fitting the data. Since $\delta_{\mu}$ reduces the information of the whole population \(F\) to a single point in \(\mathbb R\), the discrepancy \(\| F - F_{\delta_{\mu}}\|_p\) quantifies the largest possible error attained by a model. Therefore, the \textit{AGoF statistic}
\begin{equation}\label{Proportion-improvement}
   G(F,\mathcal{G}) = 1 - \frac{\|F-G(\btheta_F)\|_p}{\| F - F_{\delta_{\mu}}\|_p}
\end{equation}
represents the proportion of improvement of model $\mathcal{G}$ with respect to the non-informative (constant) $\delta_{\mu}$ in the approximation of $F$. Observe that, in general, $G(F,\mathcal{G})\in [0,1]$ and the extreme values $0$ and $1$ are achieved if $\mathcal{G}$ is the least informative model and $F\in \mathcal{G}$, respectively. Thus, a high value of the coefficient \eqref{Proportion-improvement} would indicate a good fit while a low value would amount to a poor approximation.

\section{Processes with estimated parameters} \label{Section.AsympDistr}

We present the theoretical results that guarantee the validity of the proposed methodology.

\subsection{Asymptotic behaviour in \(\mathrm{L}^p\)}

To establish the asymptotic result \eqref{WeakLimit}, first we obtain the weak limit in the space \(\mathrm{L}^p\) of the underlying  process
\begin{equation}\label{Process-Gn(theta)}
\mathbb{G}_n({\boldtheta}_F)=\sqrt{n}( \bbF_n- F) -  \sqrt{n}(  G(\hat\boldtheta_n)- G({\boldtheta}_F)).
\end{equation}

When $F=G({\boldtheta}_F)\in \mathcal{G}$, then $\mathbb{G}_n({\boldtheta}_F)=\sqrt{n}(  \bbF_n- G(\hat\boldtheta_n))$ is the \textit{empirical process with estimated parameters}.
Conversely, if $F\notin \mathcal{G}$ then the parameter $\boldtheta$ is estimated with data coming from a distribution not belonging to the family $\mathcal{G}$, in which case the use of M-estimators facilitates the analysis.

The first assumption to deal with the process \(\mathbb{G}_n({\boldtheta}_F)\) 
is that the function $G(\boldtheta)$ depends on $\boldtheta$ in a smooth way around $\boldtheta_F$ with respect to the $\mathrm{L}^p$-norm. 

\textbf{Assumption 1:}  The map $\boldtheta \mapsto G(\boldtheta)$ ($\btheta\in \Theta\subset \R^k$) satisfies that there exists a function $\dot{\G}(\boldtheta_F):\R\to \R^k$, with components $\dot{G}_1(\boldtheta_F),\dots,\dot{G}_k(\boldtheta_F)\in \mathrm{L}^p(\R)$, such that
\begin{equation} \label{Frechet}
\Vert G({\boldtheta}_F + \mathbf{h})- G({\boldtheta}_F)-\dot{\G}({\boldtheta}_F)^T  \mathbf{h}\Vert_p = o(\Vert \mathbf{h}\Vert), \quad \mathbf{h}\to \boldsymbol{0},
\end{equation}
where $\Vert \cdot \Vert$ is the Euclidean norm in $\R^k$.

Condition (\ref{Frechet}) is usually satisfied in all the examples in which $G(x;\boldtheta)$ is a smooth function of $\boldtheta\in\Theta$ and is fulfilled by many important parametric families of distributions.



The second assumption is related to the sequence of estimators $\{ \hat{\boldsymbol{\theta}}_n \}$.

\textbf{Assumption 2:} The map $\boldsymbol{\theta}\mapsto \E_F \boldsymbol{\psi}_{\boldsymbol{\theta}}(X)$ is differentiable at $\boldsymbol{\theta}_F$ with non-singular $(k\times k)$ derivative matrix $\boldsymbol{V}_{\boldsymbol{\theta}_F}$. Additionally, $\E_F\Vert \boldsymbol{\psi}_{\boldsymbol{\theta}_F}(X)\Vert^2<\infty$ and
\begin{equation}\label{asymptotically-linear}
\sqrt{n}(\hat{\boldsymbol{\theta}}_n-\boldsymbol{\theta}_F)=-\boldsymbol{V}_{\boldsymbol{\theta}_F}^{-1}\frac{1}{\sqrt{n}}\sum_{i=1}^n \boldsymbol{\psi}_{\boldsymbol{\theta}_F}(X_i)+o_\P(1).
\end{equation}

Assumption~2 requires that the M-estimator $\hat{\boldsymbol{\theta}}_n$ admits an
\emph{asymptotic linear representation}, a standard property in the asymptotic theory of
M-estimators; see \cite[Theorem~5.23]{van der Vaart} and \cite{Lehmann-Casella}.
It holds under mild regularity conditions such as differentiability of the estimating
function, identifiability of the parameter, and the existence of a finite non-singular
Fisher information matrix. Examples include maximum likelihood estimators for classical
parametric families (normal, exponential, gamma, logistic) and standard method-of-moments
estimators for location and scale parameters; see \cite[Section~7]{Serfling-80}.

We also need to impose some integrability condition on $X$, i.e., on the cdf $F$.
Specifically, we assume that $X\in\mathcal{L}^{2/p,1}$, the Lorentz space of r.v. such that
\begin{equation}\label{Lambda}
\int_0^\infty {\P(|X|>t)}^{p/2}\,  \d t < \infty.
\end{equation}

The parameter $p$ serves to modulate the weight of the tails in the model validation. Small $p$-s generate discrepancies in which the tails have a greater relevance. This is also noticeable in the condition $X\in\mathcal{L}^{2/p,1}$. For $p>0$, we denote by $\mathcal{L}^p$ the space of r.v. $X$ with finite $p$-th moment, that is, $\E |X|^p<\infty$. It can be checked (see \cite{Grafakos}) that if $1\le p<2$, then $\mathcal{L}^{2/p,1}\subset \mathcal{L}^{2/p}$. In particular, $X\in\mathcal{L}^{2,1}$ is slightly stronger than $\E X^2<\infty$ (second finite moment). For $p=2$, $\mathcal{L}^{1,1}=\mathcal{L}^1$, the space of integrable r.v. ($X$ such that $\E |X|<\infty$). For $2< p<\infty$, $\mathcal{L}^{2/p}\subset \mathcal{L}^{2/p,1}$, that is, \eqref{Lambda} is weaker than $\E |X|^{2/p}<\infty$. Hence, condition $X\in\mathcal{L}^{2/p,1}$ is more demanding when $p$ is small and relaxes as $p$ gets larger.

Theorem~\ref{Theorem.Process.parameters} is the building block for the weak convergence in \eqref{WeakLimit}, needed to derive a rejection region for the AGoF test \eqref{AGoFspecific}.
We recall that if $\mathbb{S}_n$ and $\mathbb{S}$ are stochastic processes with trajectories in
$\mathrm{L}^p$, then weak convergence $\mathbb{S}_n \to_w \mathbb{S}$ in $\mathrm{L}^p$
means that $ \E f(\mathbb{S}_n) \to \E f(\mathbb{S})$, for every continuous and bounded functional $f:\mathrm{L}^p \to \mathbb{R}$.


\begin{theorem}\label{Theorem.Process.parameters}
Let Assumptions 1 and 2 hold. Denote by $\B$ the standard Brownian bridge on $[0,1]$ and by $\B_F=\B\circ F$ the \textit{$F$-Brownian bridge}. The following two conditions are equivalent:
\begin{enumerate}[label=(\roman*), topsep=0mm, itemsep=1mm, parsep=0mm, align=left, labelwidth=*, leftmargin=-0.5 mm]
\item $X\in\mathcal{L}^{2/p,1}$.
\item $\mathbb{G}_n({\boldtheta}_F)\cw \mathbb{G}_{{\boldtheta}_F}$ in  $\mathrm{L}^p$, where $\mathbb{G}_{{\boldtheta}_F}$ is a centered Gaussian process with continuous paths a.s. and covariance function given by
\begin{equation*}
\begin{split}
\Cov(\mathbb{G}_{{\boldtheta}_F}(x),\mathbb{G}_{{\boldtheta}_F}(y))=&F(x\wedge y)-F(x)F(y)+ \dot{\G}(x,{\boldtheta}_F)   \mathbf{M}_{{\boldsymbol{\theta}_F}}  \dot{\G}(y,{\boldtheta}_F)^T  \\
&- \dot{\G}(x,{\boldtheta}_F)^T   \E_F\left[ \mathbf{l}_{{\boldsymbol{\theta}_F}}(X) 1_{\{X\le y  \}} \right]\\
&- \dot{\G}(y,{\boldtheta}_F)^T   \E_F\left[ \mathbf{l}_{{\boldsymbol{\theta}_F}}(X) 1_{\{X\le x  \}} \right],
\end{split}
\end{equation*}
for all $x,y\in\R$, where $\mathbf{l}_{{\boldsymbol{\theta}_F}}=- \boldsymbol{V}_{\boldsymbol{\theta}_F}^{-1} \boldsymbol{\psi}_{\boldsymbol{\theta}_F}$ is the influence function in (\ref{asymptotically-linear})
with covariance matrix
\begin{equation}\label{matrix M}
\mathbf{M}_{{\boldsymbol{\theta}_F}}\equiv \E_F\left[ \mathbf{l}_{{\boldsymbol{\theta}_F}}(X) \mathbf{l}_{{\boldsymbol{\theta}_F}}(X)^T \right]=
\boldsymbol{V}_{\boldsymbol{\theta}_F}^{-1} \; \E_F \left[    \boldsymbol{\psi}_{\boldsymbol{\theta}_F}(X) \boldsymbol{\psi}_{\boldsymbol{\theta}_F}(X)^T   \right] \, (\boldsymbol{V}_{\boldsymbol{\theta}_F}^{-1})^T.
\end{equation}
\end{enumerate}
\end{theorem}


We now derive the asymptotic distribution \eqref{WeakLimit} of the \(\mathrm{L}^p\)-distance between the empirical distribution and the estimated parametric model.
The proof of this result follows from Theorem~\ref{Theorem.Process.parameters}, together with an extended version of the functional delta method for Hadamard directionally differentiable functionals (see \cite{Fang-Santos}). We note that the continuous mapping theorem can only be applied when $F = G(\boldsymbol{\theta}_F)$, i.e., when $\| F - G(\boldsymbol{\theta}_F) \|_p = 0$, which corresponds to the usual null hypothesis in classical goodness-of-fit tests. However, in the general setting of the AGoF test, where, under $H_0$, $\| F - G(\boldsymbol{\theta}_F) \|_p > 0$, it is necessary to apply the delta method, which requires Hadamard (directional) differentiability.


\begin{theorem}\label{Theorem-probability-metric-parameters}
Let Assumptions 1 and 2 be satisfied and let $\mathbb{G}_{{\boldtheta}_F}$ be the process in Theorem~\ref{Theorem.Process.parameters} (ii). If $X\in \mathcal{L}^{2/p, 1}$, then the weak convergence in \eqref{WeakLimit} holds with the following asymptotic distributions.
\begin{enumerate}[label=(\alph*), topsep=0mm, itemsep=1mm, parsep=0mm, align=left, labelwidth=*, leftmargin=-0.5 mm]
\item When $p=1$,
\begin{equation} \label{AsympDistr_p_equal_1}
T(F,G({\boldtheta}_F),1)=\int_{C_{{\boldtheta}_F}} |  \mathbb{G}_{{\boldtheta}_F}  | + \int_{\R\setminus {C_{{\boldtheta}_F}}} \mathbb{G}_{{\boldtheta}_F} \, \sgn(F-G({\boldtheta}_F)),
\end{equation}
where $C_{{\boldtheta}_F}=\{t\in\R : F(t)=G({\boldtheta}_F;t)\}$ is the contact set of $F$ and $G({\boldtheta}_F)$ and $\sgn(\cdot)$ is the sign function.
\item When $1<p<\infty$, if $F=G({\boldtheta}_F)$ then $T(F,G({\boldtheta}_F),p)=\Vert  \mathbb{G}_{{\boldtheta}_F}  \Vert_p$, and if $F\ne G({\boldtheta}_F)$ then
\begin{equation} \label{AsympDistr_p_larger_1}
T(F,G({\boldtheta}_F),p)=\frac{1}{\Vert F-G({\boldtheta}_F)\Vert_p^{p-1}}\int  \mathbb{G}_{{\boldtheta}_F} \, |F-G({\boldtheta}_F)|^{p-1}\,\sgn(F-G({\boldtheta}_F)).
\end{equation}
\end{enumerate}
\end{theorem}

The following corollary specifies necessary and sufficient conditions for the limit variable $T(F,G({\boldtheta}_F),p)$ in Theorem \ref{Theorem-probability-metric-parameters} to be normal. This is useful when computing the critical value \eqref{c_alpha} in the rejection region of the AGoF test (see Section~\ref{Subsection.Bootstrap}).

\begin{corollary}\label{Corollary normalidad}
Under the conditions of Theorem \ref{Theorem-probability-metric-parameters}, we have that
\begin{enumerate}[label=(\roman*), topsep=0mm, itemsep=1mm, parsep=0mm, align=left, labelwidth=*, leftmargin=-0.5 mm]
\item If $p=1$, $T(F,G({\boldtheta}_F),1)$ has zero mean normal distribution if and only if the Lebesgue measure of the contact set $C_{{\boldtheta}_F}=\{F=G({\boldtheta}_F)\}$ is zero.

\item If $1<p<\infty$, $T(F,G({\boldtheta}_F),p)$ has zero mean normal distribution if and only if
$F\ne G({\boldtheta}_F)$, that is, whenever $F$ does not belong to $\mathcal{G}$.
\end{enumerate}
\end{corollary}

\subsection{Bootstrap consistency} \label{Subsection.Bootstrap}

The rejection region of the AGoF test is determined by \(c_n(\alpha)\) in \eqref{c_alpha}, which depends on a quantile of the limit  \(T(F,G(\btheta_F),p)\) in Theorem \ref{Theorem-probability-metric-parameters}. However, the latter depends on an integral of the stochastic process \(\mathbb{G}_{{\boldtheta}_F}\), which in turn has a complicated expression for the covariance function (where several terms have to be estimated). Here, we propose a simpler bootstrap-based procedure to approximate the quantiles of \(T(F,G(\btheta_F),p)\).

First, we have to prove the bootstrap consistency, that is, that the limit distribution of  \(T_{n}(F,G({\boldtheta}_F),p)\) in \eqref{WeakLimit} and that of its bootstrap version coincide with probability~1.
Let \(X_1^*,\ldots,X_n^*\) be a standard bootstrap sample from \(\bbF_n\), denote by \(\bbF_n^*\) its empirical distribution and by \(\hat\btheta_n^*\) the solution of the equations
\[
\int \bpsi_{\btheta} (x) \,\d \bbF_n^*(x) = \frac{1}{n} \sum_{i=1}^n \bpsi_{\btheta}(X_i^*) = \bcero.
\]
The bootstrap version of the process in \eqref{Process-Gn(theta)} is
\begin{equation} \label{EmpProcEstParBOOT}
\bbG_n^*(\hat\btheta_n) = \sqrt{n} (\bbF_n^*-\bbF_n) - \sqrt{n}(G(\hat\btheta_n^*)-G(\hat\btheta_n)).
\end{equation}

To establish the consistency of this bootstrap process, we introduce some extra assumptions.

{\bf Assumption 3:} The M-estimator is strongly consistent, that is, \(\hat\btheta_n \to \btheta_F\) a.s.

{\bf Assumption 4:} The bootstrap M-estimator \(\hat\btheta_n^*\) is consistent in \(\bbF_n\)-probability with \(F\)-probability 1. More formally,  for every $\epsilon > 0$,
\[
\P_{\bbF_n}\left(  \left\| \hat{\boldsymbol{\theta}}_n^* - \hat{\boldsymbol{\theta}}_n \right\| > \epsilon  \right) = \P\left( \left\| \hat{\boldsymbol{\theta}}_n^* - \hat{\boldsymbol{\theta}}_n \right\| > \epsilon \,\middle|\, X_1, \dots, X_n \right) \to 0 \quad \text{as } n \to \infty,
\]
for almost every sample $(X_1, X_2, \dots)$ drawn from $F$. Here, $\P_{\bbF_n}(\cdot) = \P(\cdot \mid X_1,\dots,X_n)$ denotes the bootstrap probability given the data, i.e., the conditional probability assuming the observations are sampled from the empirical distribution \(\bbF_n\) of $(X_1, \dots, X_n)$. In particular, in the expression above, $\hat{\boldsymbol{\theta}}_n$ is considered fixed (non-random). This convergence is often referred to as \textit{conditionally almost sure} convergence; see \cite[Chapter~23]{van der Vaart-Wellner 2023}.


{\bf Assumption 5:} It holds that
\begin{eqnarray*}
\lefteqn{\sqrt{n} ( \hat\btheta_n^* - \hat\btheta_n )
+ \bV_{\btheta_F}^{-1} \frac{1}{\sqrt{n}} \sum_{i=1}^n (\bpsi_{\btheta_F}(X_i^*)-\bpsi_{\btheta_F}(X_i)) } \\
 & & \hspace{12 mm} = \sqrt{n} ( \hat\btheta_n^* - \hat\btheta_n )
+ \bV_{\btheta_F}^{-1} \sqrt{n} (\bbF_n^*-\bbF_n)(\bpsi_{\btheta_F}) \xrightarrow[]{\P} 0 \quad F\mbox{-a.s.}
\end{eqnarray*}

\cite{Huber-67} established conditions under which Assumption 3 holds (see also \cite[Ch. 7]{Serfling-80}).
\cite[Thm. 3.7]{Arcones-Gine-92} proved that Assumption 4 is fulfilled under the same conditions used by \cite{Huber-67} to prove the \(F\)-a.s. consistency of the M-estimator \(\hat\btheta_n\) (Assumption 3). 
Assumption~5 is the bootstrap analogue of Assumption~2 and requires the bootstrap
to replicate the first-order behaviour of the estimator, thus ensuring the
\emph{asymptotic linearity} of \(\hat{\boldsymbol{\theta}}_n^*\).
\cite[Theorem~3.6]{Arcones-Gine-92} give conditions for this property and use them to prove
the asymptotic normality of the bootstrap estimator.
Moreover, \cite[Theorem~2.2 and Corollary~2.3]{Burke-Gombay}
show that Assumptions~4 and~5 hold for maximum likelihood estimators under the
standard regularity conditions needed to define and compute Fisher information.

The next theorem establishes the a.s.\ consistency of the bootstrap process
$\bbG_n^*(\hat\btheta_n)$ in $\mathrm{L}^p$,
a key step for proving the consistency of the bootstrap estimator of the test statistic.
Specifically, we show that
$\bbG_n^*(\hat\btheta_n) \cw \bbG_{\btheta_F}$ in $\mathrm{L}^p$
for almost every sample $(X_1,\ldots,X_n,\dots)$ from $F$,
that is, $\E_{\mathbb{F}_n} f(\mathbb{G}_n^*\hat({\boldsymbol{\theta}}_n))
\to \E f(\mathbb{G}_{\boldsymbol{\theta}_F})$, as $n\to\infty$,
for all continuous and bounded $f:\mathrm{L}^p\to \mathbb{R}$.

\begin{theorem} \label{Theorem.Bootstrap.Consistency}
Let Assumptions 1--5 hold. For $1\le p<2 $, let us assume that \(X\in\mathcal L^{2/p,1}\) and for $2\le p<\infty $ that \(X\in\mathcal L^{2/p}\). Then, the bootstrap process \(\bbG_n^*(\hat\btheta_n)\) in \eqref{EmpProcEstParBOOT} is consistent in \(\mathrm{L}^p\) with probability 1.
\end{theorem}

Thanks to the differentiability of the $\mathrm L^{p}$-norm, together with Theorem~\ref{Theorem.Bootstrap.Consistency} and \cite[Thm. 3.1]{Fang-Santos}, we conclude the desired consistency of bootstrap test statistic
\begin{equation} \label{Bootstrap_E_n}
T_{n}^*(\bbF_n,G(\hat\btheta_n),p) = \sqrt{n}(\|\bbF_n^*-G(\hat\btheta_n^*)\|_p-\|\bbF_n-G(\hat\btheta_n)\|_p).
\end{equation}
\begin{corollary} \label{Corollary.Bootstrap.Consistency.Lp}
Under the assumptions of Theorem \ref{Theorem.Bootstrap.Consistency}, let us further assume that for $p=1$ the contact set $C_{{\boldtheta}_F}$ in Theorem \ref{Theorem-probability-metric-parameters} has zero measure. Then, the statistic \eqref{Bootstrap_E_n} converges weakly to \(T(F,G(\btheta_F),p)\), the limit distribution in Theorem~\ref{Theorem-probability-metric-parameters}, with probability 1.

\end{corollary}

\subsection{Practical implementation} \label{Subsection.Implementation}

We have applied the bootstrap procedure in two ways (asymptotically equivalent).
The first option is to note that
\begin{eqnarray*}
\alpha & \simeq & \P_{\bbF_n} \left\{ \sqrt{n}(\|\bbF_n^*-G(\hat\btheta_n^*)\|_p - \|\bbF_n-G(\hat\btheta_n)\|_p) \leq Q_{T(F,G(\btheta_F),p)}(\alpha) \right\} \\
 & = & \P_{\bbF_n} \left\{\|\bbF_n^*-G(\hat\btheta_n^*)\|_p \leq \|\bbF_n-G(\hat\btheta_n)\|_p - c_n(\alpha) \right\}.
\end{eqnarray*}
So \(\|\bbF_n-G(\hat\btheta_n)\|_p - c_n(\alpha)\) is approximately \(\epsilon^{*(\alpha)}\), the \(\alpha\)-quantile of \(\|\bbF_n^*-G(\hat\btheta_n^*)\|_p\), that is,
\(- c_n(\alpha) \simeq \epsilon^{*(\alpha)}-\|\bbF_n-G(\hat\btheta_n)\|_p\).
Consequently, by \eqref{c_alpha}, we reject \(H_0\) in \eqref{AGoFspecific} at a significance level \(\alpha\) when $2\|\bbF_n-G(\hat\btheta_n)\|_p - \epsilon^{*(\alpha)} < \epsilon$.

The second procedure is valid when the asymptotic distribution \(T(F,G(\btheta_F),p)\) is normal with expectation 0 and standard deviation \(\sigma_a\) (see Corollary~\ref{Corollary normalidad}). In this case, we have that, with probability 1, for \(n\) large, \(\|\bbF_n^*-G(\hat\btheta_n^*)\|_p\) follows approximately a normal distribution with expectation \( \|\bbF_n-G(\hat\btheta_n)\|_p\) and standard deviation \(\sigma_{\mbox{\scriptsize boot}}=\sigma_a/\sqrt{n}\).
Then, we reject \(H_0\) at (asymptotic) level \(\alpha\) when
$\|\bbF_n-G(\hat\btheta_n)\|_p -\sigma_{\mbox{\scriptsize boot}} z_{\alpha} < \epsilon,$
where $z_{\alpha}$ is the $\alpha$-quantile of a standard normal distribution.
We call Bootstrap 1 and 2 the methods with rejection regions obtained by these two procedures.

Specifically, given an observed sample $x_1,\dots,x_n$ from $F$, the quantile \(\epsilon^{*(\alpha)}\) and the standard deviation \(\sigma_{\mbox{\scriptsize boot}}\) have been approximated via resampling as follows.

\noindent \textbf{Step 1.} Extract \(B\) bootstrap samples from \(\bbF_n\).
\begin{equation*} 
    \begin{array}{cccc}
\mbox{Original sample} & & \mbox{Bootstrap samples} \\
\begin{array}{l}
x_{1},\ldots,x_{n}
\end{array}
&
\begin{array}{c}
\longrightarrow
\end{array}
&
\begin{array}{l}
x_{1}^{*b},\ldots,x_{n}^{*b},
\end{array}
&
b=1,\ldots,B.
\end{array}
\end{equation*}

\noindent \textbf{Step 2.} For each bootstrap sample $x_{1}^{*b},\ldots,x_{n}^{*b}$, compute its empirical cdf, $\bbF_n^{*b}$, and the corresponding M-estimator, $\hat\btheta_n^{*b}$, to obtain the approximated value \(\|\bbF_n^{*b}-G(\hat\btheta_n^{*b})\|_p\).
\begin{equation*} 
    \begin{array}{cccc}
\mbox{Bootstrap samples} & & \mbox{Bootstrapped norms} \\
\begin{array}{l}
x_{1}^{*b},\ldots,x_{n}^{*b}
\end{array}
&
\begin{array}{c}
\longrightarrow
\end{array}
&
\begin{array}{l}
\|\bbF_n^{*b}-G(\hat\btheta_n^{*b})\|_p,
\end{array}
&
b=1,\ldots,B.
\end{array}
\end{equation*}

\noindent \textbf{Step 3.} Calculate \(\hat{\epsilon}^{*(\alpha)}\), the $\alpha$-quantile, of the values \(\|\bbF_n^{*b}-G(\hat\btheta_n^{*b})\|_p\), as well as its standard deviation \(\hat{\sigma}_{\mbox{\scriptsize boot}}\).
\begin{equation*} 
    \begin{array}{cccc}
\mbox{Bootstrapped norms} & & \mbox{$\alpha$-quantile and s.d.} \\
\begin{array}{l}
\{ \|\bbF_n^{*b}-G(\hat\btheta_n^{*b})\|_p\}_{b=1}^B
\end{array}
&
\begin{array}{c}
\longrightarrow
\end{array}
&
\begin{array}{l}
\hat\epsilon^{*(\alpha)}\),\, \(\hat\sigma_{\mbox{\scriptsize boot}}.
\end{array}
&
\end{array}
\end{equation*}

\noindent \textbf{Step 4.}  Apply the Bootstrap 1 and 2 methods.
\begin{itemize}
    \item[--] Bootstrap 1: Reject \(H_0\) in \eqref{AGoFspecific} at a significance level \(\alpha\) when
\begin{equation} \label{RejectionRegBoot1}
2\|\bbF_n-G(\hat\btheta_n)\|_p - \hat\epsilon^{*(\alpha)} < \epsilon.
\end{equation}

 \item[--]  Bootstrap 2: Reject \(H_0\) in \eqref{AGoFspecific} at a significance level \(\alpha\) when
\begin{equation} \label{RejectionRegBootAsymp}\|\bbF_n-G(\hat\btheta_n)\|_p -\hat\sigma_{\mbox{\scriptsize boot}} z_{\alpha} < \epsilon.
\end{equation}
\end{itemize}



\section{A simulation study} \label{Section.Simulations}

To check the performance of the AGoF testing procedure we have carried out a simulation study with various models.
For each of 1000 Monte Carlo runs we have generated one sample of size \(n\) from \(X\) and drawn \(B=2000\) bootstrap samples to approximate \(\epsilon^{*(\alpha)}\) and \(\sigma_{\mbox{\scriptsize boot}}\). The chosen sample sizes are \(n=30, 50, 100, 500\) for each of the models under consideration. The significance level in all cases is $\alpha=0.05$. By checking whether \(\epsilon\) fulfills \eqref{RejectionRegBoot1} or \eqref{RejectionRegBootAsymp} or not, we obtain the proportion of \(H_0\) rejections for each possible value of \(\epsilon\), i.e., the power of the test.

All assumptions required for the theoretical results are satisfied by the models considered in this
simulation study. Assumption~1 holds because the model distribution functions have finite
$\mathrm{L}^p$-norms of their second derivatives in a neighbourhood of $\boldsymbol{\theta}_F$.
Assumptions~2--3 are fulfilled when the estimator is either the maximum likelihood estimator or a
method-of-moments estimator under standard regularity conditions, while Assumptions~4 and~5 for the
bootstrap estimator follow from \cite{Burke-Gombay}.
All parametric models used in the simulations (normal, exponential, gamma, etc.) meet these conditions,
as they belong to standard families with well-defined and finite Fisher information.

\subsection*{\sl A Weibull distribution and the exponential model}

We consider the exponential model,
$$\mathcal G = \{G_\theta(x)=1-e^{-x/\theta}:  x>0 ,\,  \theta>0 \}.$$
The variable \(X\) follows a Weibull distribution with shape parameter \(2\) and scale parameter 1, that is, \( F(x) = 1 - e^{-x^2} \), \(x>0\). We select $p=1$ to better detect differences in the right tail. In this example, we have that $\theta_F=\E X = \mu=\sqrt{\pi}/2$ and the \(\mathrm{L}^1\)-distance is \(\|F-G(\theta_F)\|_1=0.3002\). The AGoF statistic in \eqref{Proportion-improvement} is \(G(F,\mathcal{G}))=0.194\). This means that the exponential model only improves by $19.4\%$ over the degenerate distribution at $\mu$ in the approximation of this Weibull variable (with respect to the $\mathrm{L}^1$-norm).
In Figure~\ref{Fig.PowerSimul1}(a) we display the power attained by the procedures Bootstrap 1 \eqref{RejectionRegBoot1} (continuous lines) and Bootstrap 2 \eqref{RejectionRegBootAsymp} (dashed lines). Observe that, for \(n=100\) both power functions already adjust well to the significance level. For \(n=500\), they are almost undistinguishable. Thus, the performance of the bootstrap rejection schemes is satisfactory for a moderately large sample size \(n\). For \(n=30\) or 50 the Bootstrap 2 procedure attains the desired significance level, while the power obtained with the Bootstrap 1 method exceeds the 5\% target.

\subsection*{\sl A Gaussian mixture and the normal model}

The parametric model is normal,
\begin{equation} \label{NormalModel}
    \mathcal G = \{G_{\btheta}(x)=\Phi((x-\mu)/\sigma) : x\in\bbR,\, \btheta=(\mu,\sigma),\, \mu\in\bbR,\, \sigma>0 \},
\end{equation}
where \(\Phi\) denotes the standard normal cdf. The variable \(X\) follows a Gaussian mixture distribution with two components, \( F(\cdot)= 0.8 \Phi + 0.2 \Phi((\cdot-2)/2)\).
We consider $p=2$. The \(\mathrm{L}^2\)-distance is \(\|F-G(\btheta_F)\|_2= 0.1081\) and \(G(F,\mathcal{G})=0.805\).
In Figure~\ref{Fig.PowerSimul1}(b)
we display the power with the rejection regions \eqref{RejectionRegBoot1} and \eqref{RejectionRegBootAsymp}. In this case, the power attained with the Bootstrap 2 method is close to the nominal 5\% significance level  for all sample sizes, while the test size with the Bootstrap 1 procedure markedly exceeds \(\alpha\) except for \(n=500\).

\subsection*{\sl A negative binomial distribution and the Poisson model}

We consider the discrete Poisson model with probability mass function
$$
\mathcal G = \left\{\P_\theta(x)=e^{-\theta} \frac{\theta^x}{x!}:  x=0,1,2\ldots ,\,  \theta>0 \right\}.
$$
The variable \(X\) follows a negative binomial distribution with parameters 3 and 2/3.
The chosen value of \(p\) is 1, for which \(\|F-G(\theta_F)\|_1=0.1793\) and \(G(F,\mathcal{G})=0.849\).
The power functions attained in the AGoF test to the Poisson model are displayed in Figure~\ref{Fig.PowerSimul1}(c). The results are similar to those of the previous model (Figure~\ref{Fig.PowerSimul1}(b)).

\subsection*{\sl The Kumaraswamy distribution and the beta model}

The sampling distribution is the Kumaraswamy(2,2) and the model is beta,
$$
\mathcal G = \left\{G_{\btheta}(x)=\frac{\Gamma(\alpha)\Gamma(\beta)}{\Gamma(\alpha+\beta)} \int_0^x t^{\alpha-1} (1-t)^{\beta-1}\,  \d t:  0<x<1 ,\,  \btheta=(\alpha,\beta) ,\, \alpha,\beta>0 \right\}.
$$
Hence, the distributions have compact support. For \(p=1\), \(\|F-G(\btheta_F)\|_1=0.0020\) and \(G(F,\mathcal{G})=0.989\). The power, displayed in Figure~\ref{Fig.PowerSimul2}(a), shows that in this case the Bootstrap 1 performs better for all the sample sizes.

\subsection*{\sl The Student \(t\) distribution and the normal model}

We consider again the normal model \eqref{NormalModel}. The sample is generated according to a \(t_4\) Student distribution. For \(p=4\), we have \(\|F-G(\theta_F)\|_4=0.0603\) and \(G(F,\mathcal{G})=0.861\). The power function appears in Figure~\ref{Fig.PowerSimul2}(b). For all sample sizes, the test size attained by the Bootstrap 2 procedure is near or below the target 5\% level, while the power of the Bootstrap 1 exceeds it.

\subsection*{\sl A lognormal distribution and the gamma model}

The gamma model is
$$
\mathcal G = \left\{G_{\btheta}(x)=\frac{\lambda^\alpha}{\Gamma(\alpha)} \int_0^x t^{\alpha-1} e^{-\lambda t}\, \d t:  x>0 ,\,  \btheta=(\alpha,\lambda) ,\, \alpha,\lambda>0 \right\}.
$$
The variable \(X\) follows a lognormal distribution with parameters \(\mu=0.5\) and \(\sigma=0.5\).
We chose \(p=1\), for which \(\|F-G(\theta_F)\|_1=0.0759\) and \(G(F,\mathcal{G})=0.897\). The power function is given in Figure~\ref{Fig.PowerSimul2}(c). As with the beta model (Figure~\ref{Fig.PowerSimul2}(a)), the Bootstrap 1 procedure performs best in this case.


\begin{figure}
\begin{center}
\begin{tabular}{c}
\includegraphics[width=0.75\textwidth]{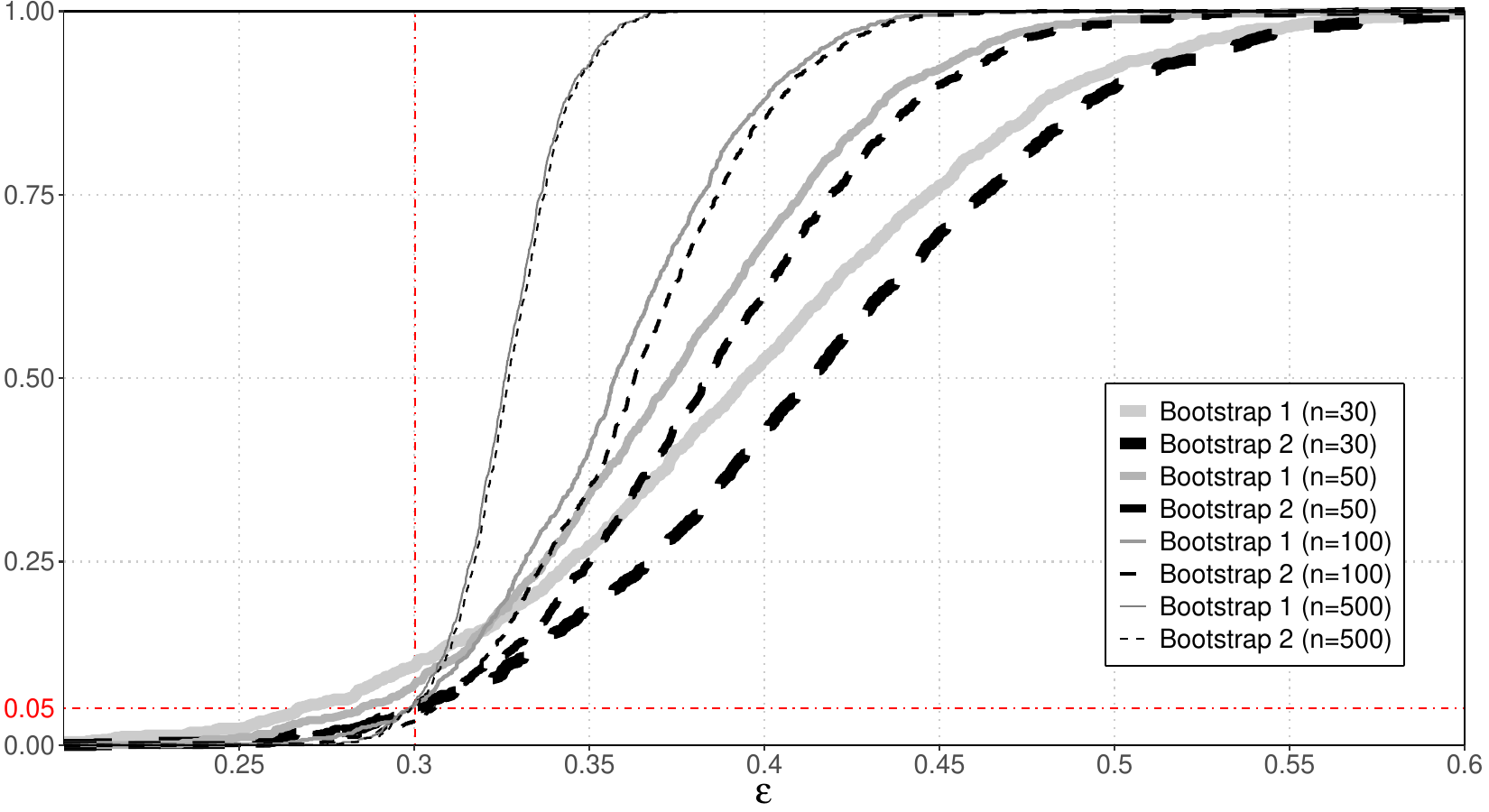} \\ (a) \\ [2 mm]
\includegraphics[width=0.75\textwidth]{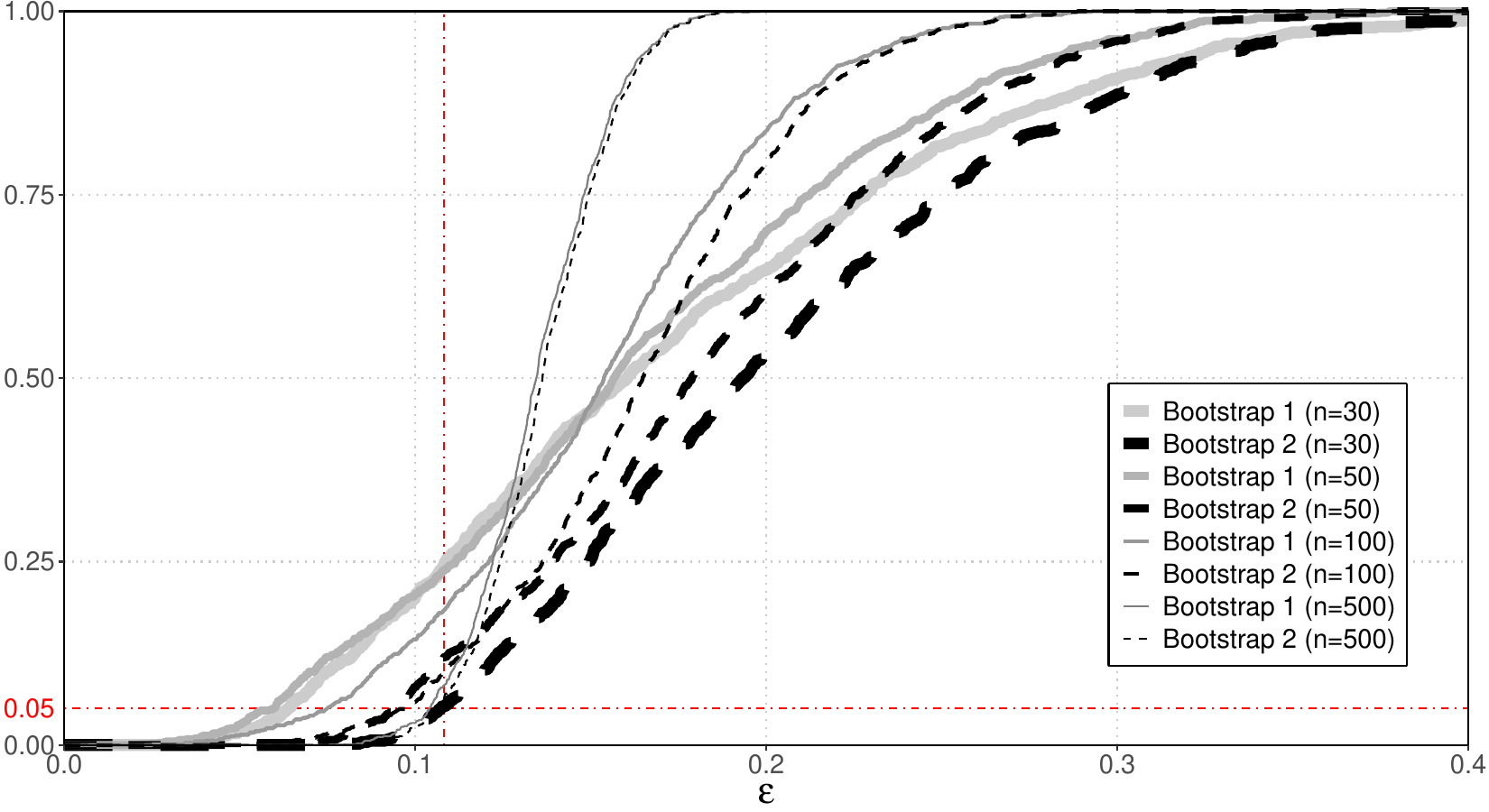} \\ (b) \\ [2 mm]
\includegraphics[width=0.75\textwidth]{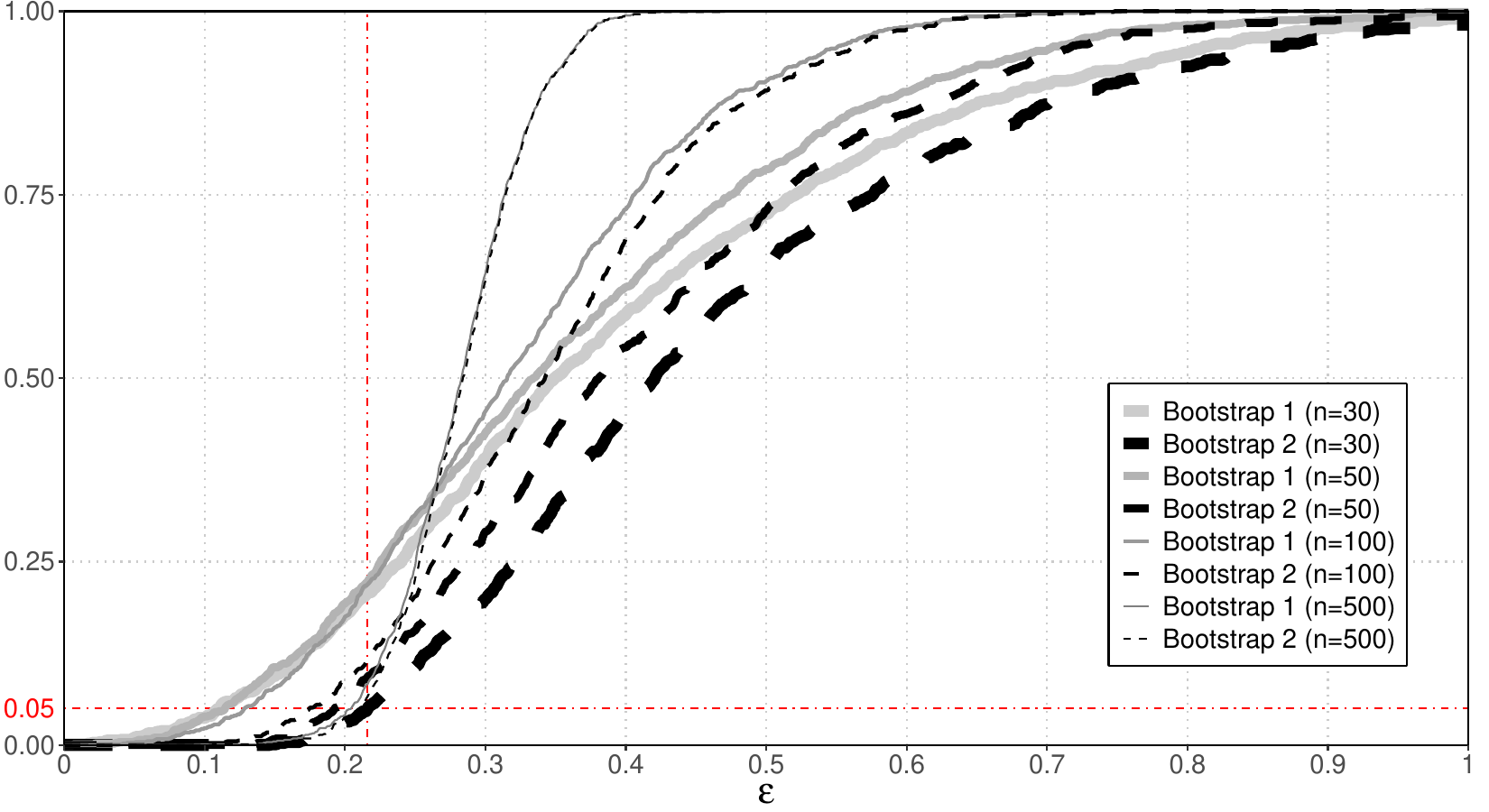} \\ (c)
\end{tabular}
\end{center}
\caption{Power function for (a) the Weibull(2,1) and the exponential model;
(b) a normal mixture and the normal model and (c) a negative binomial and a Poisson model. The vertical red line is located at \(\|F-G(\btheta_F)\|_p\).}
\label{Fig.PowerSimul1}
\end{figure}

\begin{figure}
\begin{center}
\begin{tabular}{c}
\includegraphics[width=0.75\textwidth]{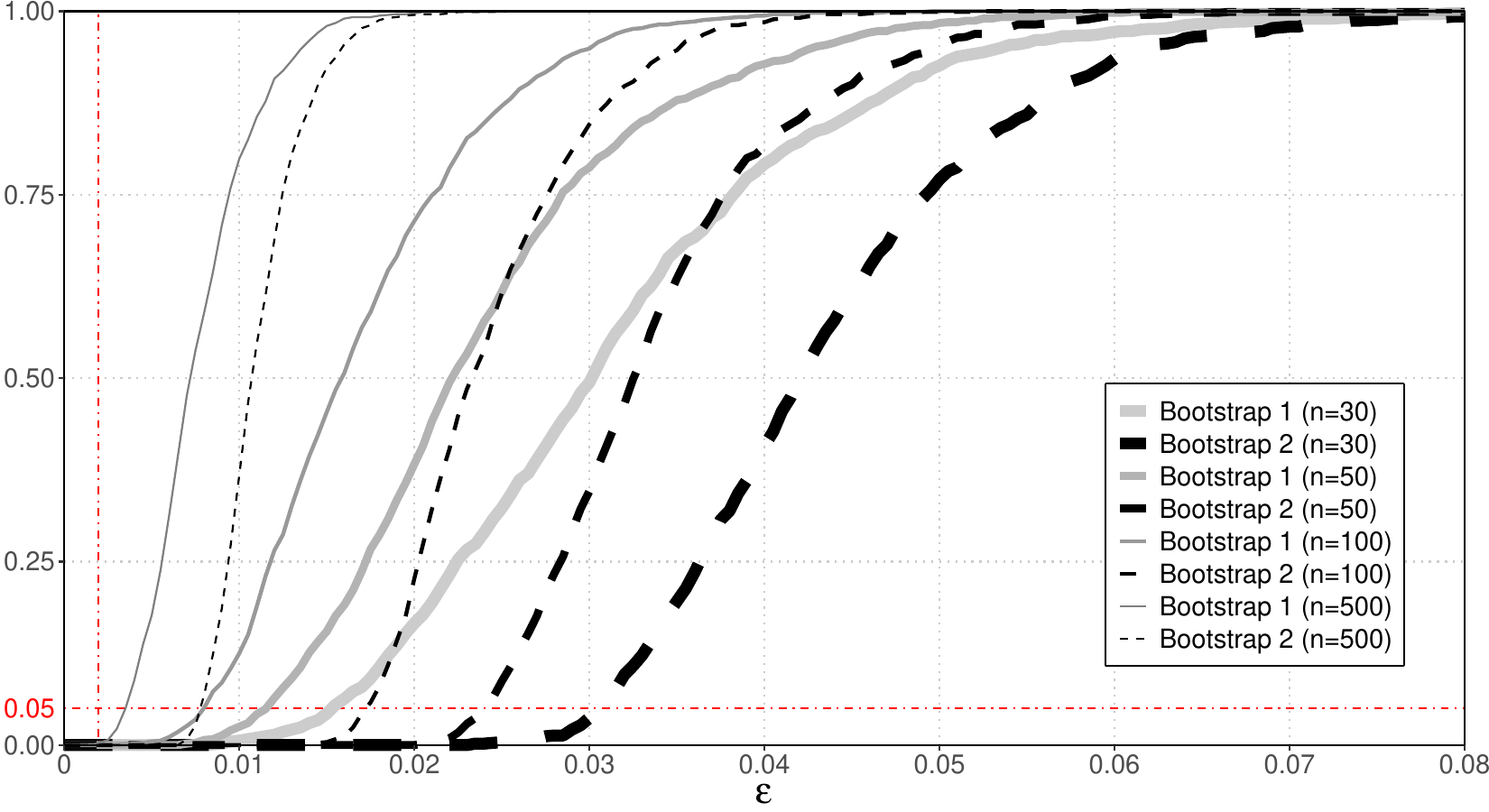} \\ (a) \\ [2 mm]
\includegraphics[width=0.75\textwidth]{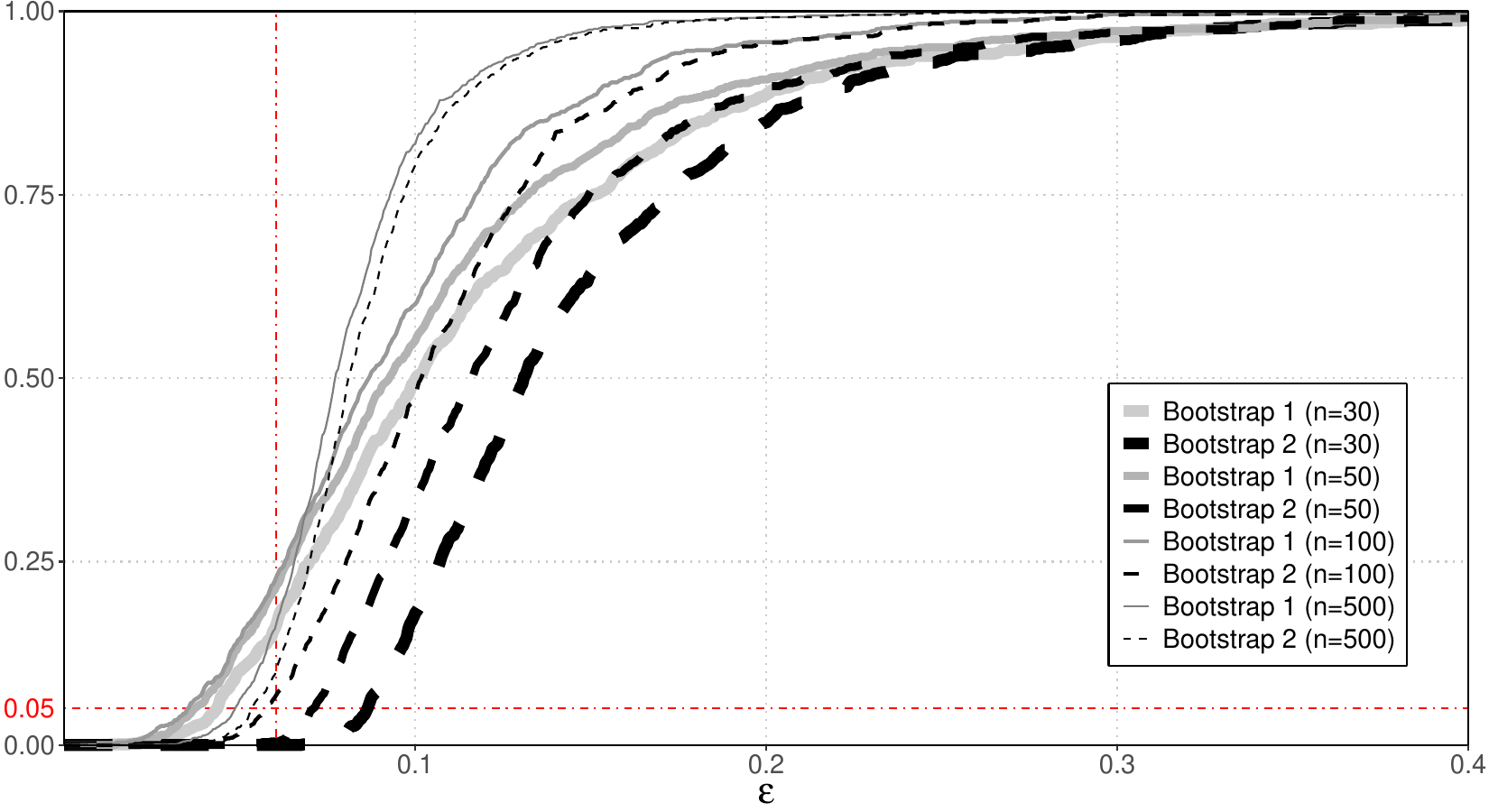} \\ (b) \\ [2 mm]
\includegraphics[width=0.75\textwidth]{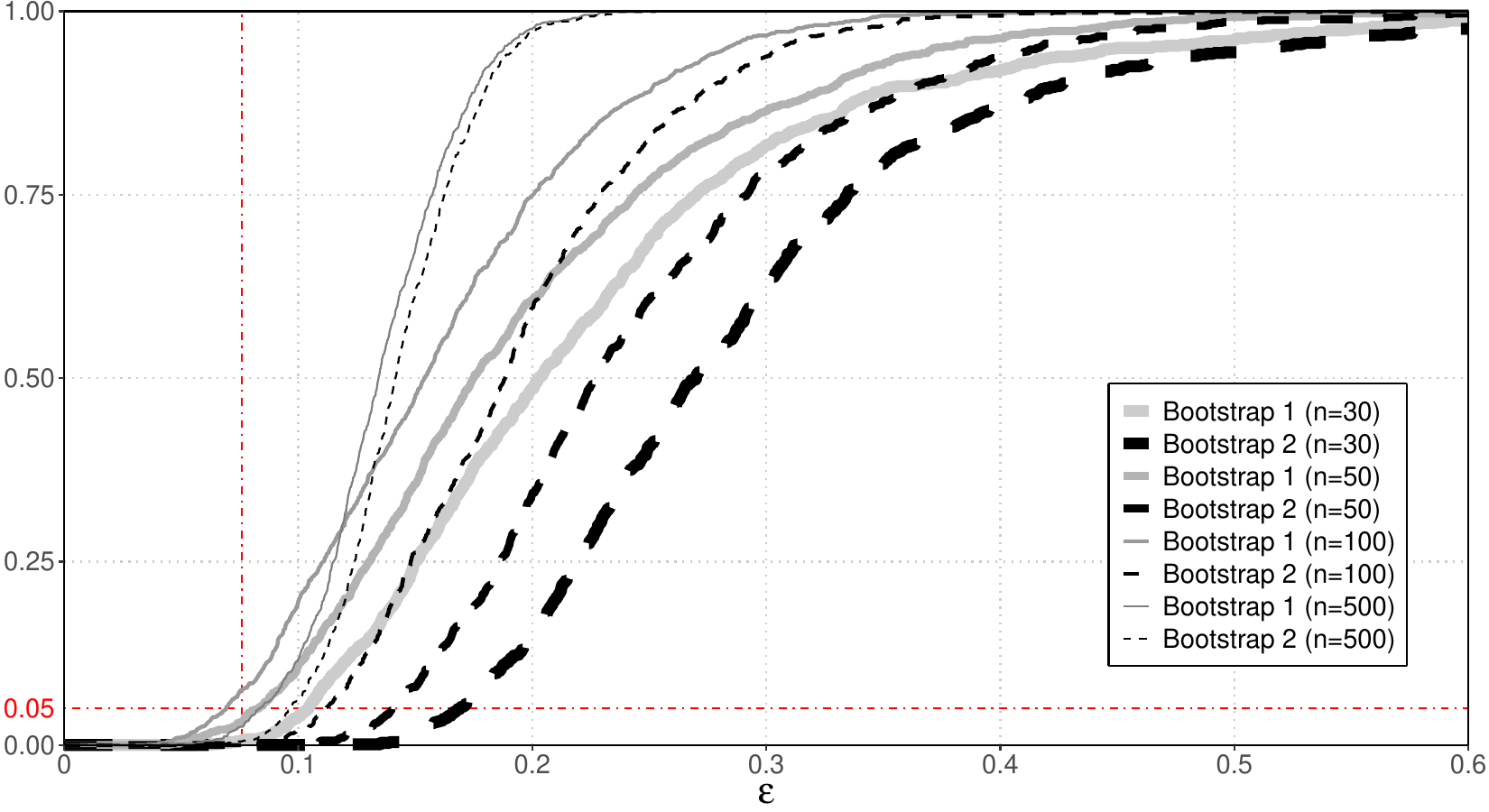} \\ (c) \\
\end{tabular}
\end{center}
\caption{Power function for (a) the Kumaraswamy(2,2) and the beta model; (b) the Student \(t_4\) and the normal model; (c) the lognormal(0.5,0.5) and the gamma model. The vertical red line is located at \(\|F-G(\btheta_F)\|_p\).}
\label{Fig.PowerSimul2}
\end{figure}

\subsection*{\sl Practical conclusions and recommendations}

Some (preliminary) practical conclusions can be drawn from the simulation study. In general terms, with intermediate sample sizes (such as $n=500$) the two proposed methods work reasonably well and attain a similar power. The size of the test is satisfactorily controlled and the power is high when the underlying distribution deviates from the null. We note that the bootstrap does not always maintain the nominal level in small samples, which is consistent with the asymptotic nature of the theoretical guarantees; see Proposition~\ref{Proposition.Properties}.

For small sample sizes ($n\le 100$), the worst results are apparently obtained for the beta model (Figure~\ref{Fig.PowerSimul2}(a)). In this example, the value of the AGoF statistic is very high ($98.9\%$) and the distance between the distributions is extremely small ($0.002$). The Kumaraswamy distribution is almost indistinguishable from its representative within the beta family. For this reason, the Gaussian approximation (Corollary \ref{Corollary normalidad}) used in Bootstrap 2 does not provide such good results. 
When the reference distribution is not so close to the model (as in Figures~\ref{Fig.PowerSimul1} (a), (b) and (c) and Figure~\ref{Fig.PowerSimul2}(b)), the significance level is usually better controlled with Bootstrap 2. Considering these empirical results, we recommend using Bootstrap 2 when the AGoF statistic in \eqref{Proportion-improvement} is not very high (values between $0$ and $0.9$) and Bootstrap 1 when the sampling distribution is very close to the model (AGoF statistic above $0.9$).


\section{Application to two real data sets}\label{Section.Data}

\subsection{IgG antibodies in Haiti serosurvey} \label{Section.AntibodiesHaiti}

\begin{figure}[h!]
\begin{center}
\includegraphics[width=0.45\textwidth]{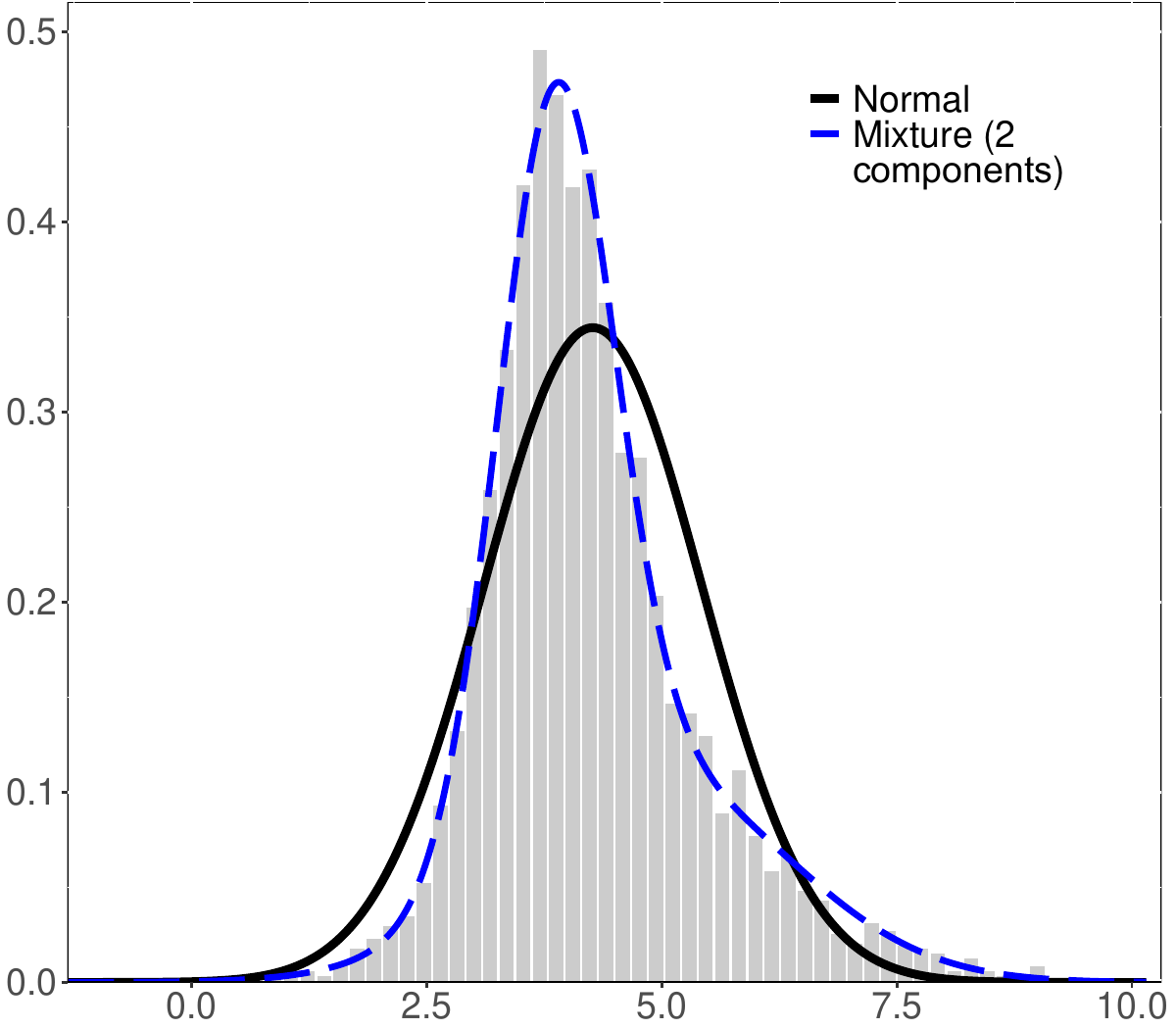}
\end{center}
\caption{Histogram of log(MFI-bg) for antigen Bm33, normal fit and 2-component normal mixture fit.}
\label{Fig.Bm33_Hist_ECDF}
\end{figure}

From December 2014 to February 2015 a nationwide serosurvey took place in Haiti. Blood samples collected from the participants were analyzed for IgG antibodies corresponding to different antigens from various pathogens (see \cite{Haiti-paper-2022}). For each antigen and participant the median fluorescence intensity minus background (MFI-bg) signal was measured. The variable of interest, \(X\) with cdf $F$, is the logarithm of the MFI-bg signal. To account for the seropositive and seronegative populations, \cite{Haiti-paper-2022} modelled the probability distribution of \(X\) for each antigen as a two-component normal mixture model.
Thus, it is interesting to check whether the latter is an appropriate model for the data.
For several antigens, we have tested the AGoF to a normal mixture distribution with \(k=1,\ldots,5\) components to analyze which of the models fits the sample best (relative to its complexity).
As the results are similar for all the antigens, we have chosen antigen Bm33 (corresponding to the pathogen {\em Lymphatic filariasis}) to illustrate the AGoF procedure.
After eliminating the missing data and the negative signals, we obtained a sample size of \(n=4308\). Figure~\ref{Fig.Bm33_Hist_ECDF} displays the histogram and the densities of a normal distribution and a 2-component normal mixture with parameters estimated by ML.


\begin{figure}[h!]
\begin{center}
\begin{tabular}{cc}
\includegraphics[width=0.46\textwidth]{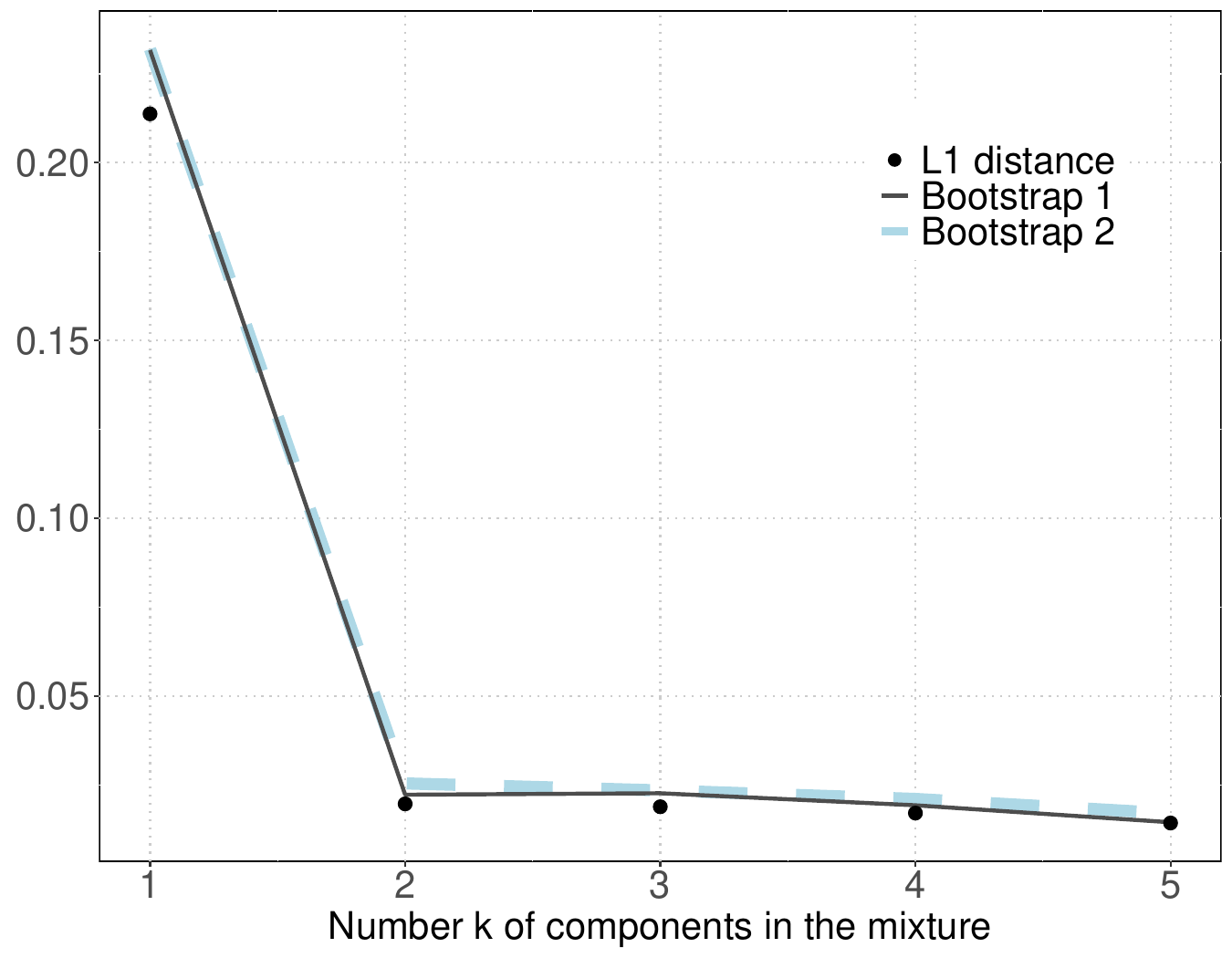} &
\includegraphics[width=0.46\textwidth]{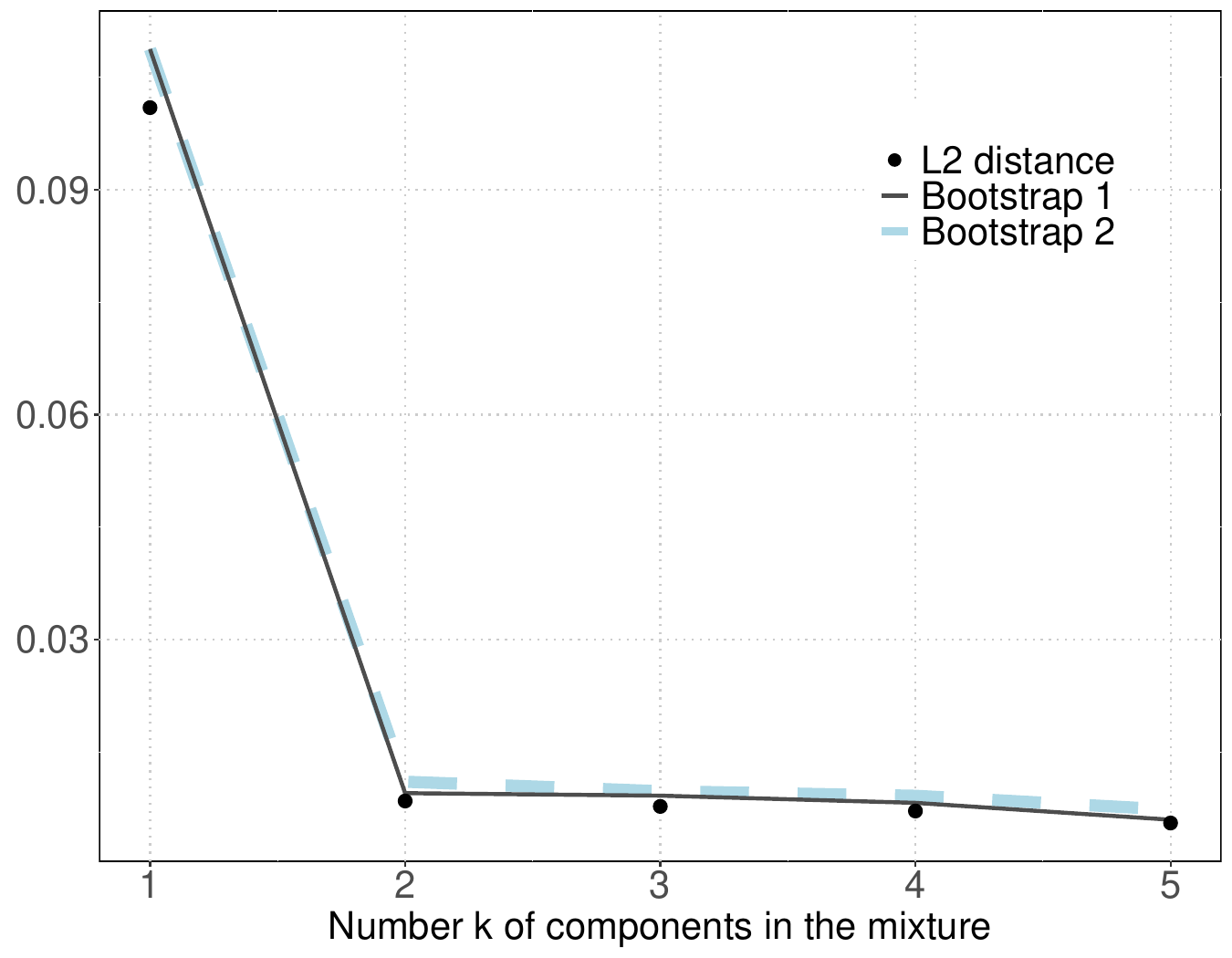} \\
(a) & (b)
\end{tabular}
\end{center}
\caption{For antigen Bm33, values of \(\epsilon_k^*(0.05)\) when (a) \(p=1\) and (b) \(p=2\). Black points are the empirical \(\mathrm{L}^p\)-distances.}
\label{Fig.Bm33_epsilon}
\end{figure}

For each number \(k\in\{1,\ldots,5\}\) of components in the mixture, we have determined the value \(\epsilon_k^*(0.05) \) (as in \eqref{minimum distance}) for which, at the significance level \(\alpha=0.05\), we reject the null hypothesis in \eqref{AGoFspecific},
where the parametric model is \({\mathcal G}_k\), the family of \(k\)-component normal mixtures.
The value of \(\epsilon_k^*\) is determined by means of the two bootstrap procedures described in Section~\ref{Subsection.Implementation} and reported in Table~\ref{Tab.Bm33_L1} for the \(\mathrm{L}^1\) and \(\mathrm{L}^2\) distances. In Figure~\ref{Fig.Bm33_epsilon} we display the \(\epsilon_k^*(0.05) \) values against \(k\). Clearly, the 2-component mixture model is the one that best fits the data with the smallest number of components. We conclude that there is \(\epsilon\)-almost goodness of fit of the log(MFI-bg) to the 2-component Gaussian mixture with \(0.22<\epsilon<0.23\) in the case of the \(\mathrm{L}^1\)-norm and with \(\epsilon\simeq 0.01\) in the case of the \(\mathrm{L}^2\)-norm.

An important issue is to interpret the magnitude of \(\epsilon\) for which we accept the AGoF
alternative hypothesis. Especially in the case of the \(\mathrm{L}^1\)-norm, values around $0.22$ may seem very large if we do not have a reference value for comparison. In the antibodies example, the situation is facilitated by the fact that the aim was to choose between nested models.
In the general case, to gain more intuition into the mentioned \(\epsilon\), we can estimate the coefficient $G(F,\mathcal{G}_k)$ defined in \eqref{Proportion-improvement}. This value represents that proportion of improvement of the model $\mathcal{G}_k$ with respect to the non-informative one given by $\delta_\mu$, with $\mu=\E(X)$. Specifically, we have computed
$${G}^*(F,\mathcal{G}_k) = 1 - \frac{\epsilon_k^*(0.05)}{\| \bbF_n - F_{\delta_{\bar{x}}}\|_p},\quad p=1,2,$$
where $\bar x$ is the sample mean. 

In the case of the log(MFI-bg) data for antigen Bm33, we obtain \(\bar x=4.2645\), \(\|\bbF_n-F_{\delta_{\bar x}}\|_1=0.8631\) and \(\|\bbF_n-F_{\delta_{\bar x}}\|_2=0.4930\). In Table~\ref{Tab.Bm33_L1} we also include \({G}^*(F,\mathcal{G}_k)\).
Observe that, for the normal model (\(k=1\)), the improvement over a constant model is less that 78\%.
This indicates that the normal distribution is not a satisfactory model for the data. When \(k\geq 2\) components are included in the Gaussian mixture, then this fraction increases up to more than 97\%. The models with $k\ge 3$ components fail to improve this percentage by more than 1\%, which reinforces the choice of the 2-component mixture model as a good approximation to the distribution generating the sample.





\begin{table}[h!]
\begin{center}
\begin{tabular}{ccccc}
 & \multicolumn{2}{c}{\(\mathrm{L}^1\)-distance} & \multicolumn{2}{c}{\(\mathrm{L}^2\)-distance} \\
\(k\) & Bootstrap 1 & Bootstrap 2 & Bootstrap 1 & Bootstrap 2 \\ \hline
1 & 0.2317 (0.732) & 0.2320 (0.731) & 0.1088 (0.779) & 0.1088 (0.779) \\
2 & 0.0222 (0.974) & 0.0254 (0.972) & 0.0095 (0.981) & 0.0110 (0.978) \\
3 & 0.0226 (0.974) & 0.0234 (0.973) & 0.0092 (0.981) & 0.0098 (0.980) \\
4 & 0.0192 (0.978) & 0.0210 (0.976) & 0.0082 (0.983) & 0.0091 (0.981) \\
5 & 0.0145 (0.983) & 0.0170 (0.980) & 0.0059 (0.988) & 0.0073 (0.985) \\ \hline
\end{tabular}
\end{center}
\caption{For antigen Bm33, values of \(\epsilon_k^*(0.05)\) and, between parentheses, \({G}^*(F,\mathcal{G}_k)\).}
\label{Tab.Bm33_L1}
\end{table}


\subsection{Failure stress of carbon fibers} \label{Section.FailureStress}

\cite{Failure-Stress-Carbon-Fibers} report tensile properties of about 1200 single carbon fibres evaluated at gauge lengths 20, 30, 40, 45, 50, 60, 65 and 80 mm, with around 150 fibres for each length (see Table~\ref{Tab.FailureStress3}). As this type of data has a slight negative skewness, these authors used Weibull distributions as a model for the failure stress of each fibre (see Figure 1 in the Supplementary Material). It is interesting to check whether the degree of almost goodness-of-fit of this model to the data varies with the gauge length. To make our analysis more complete, apart from the Weibull (W) model we have also considered the three-parameter Weibull (3W), the skew normal (SN) and a mixture of two Weibulls (the so-called bimodal Weibull, BW) as potential fits for the failure stress data. For each of these parametric models, \(\mathcal{G}\), and for each gauge length, we have computed the value \(\epsilon^*(0.05) \) and the proportion \(G^*(F,\mathcal{G})\) via the Bootstrap 1 and 2 procedures. The results appear in Table~\ref{Tab.FailureStress3} for the \(\mathrm{L}^1\) and \(\mathrm{L}^2\) metrics.
The 3-Weibull fit coincides with that of the Weibull for 7 of the 8 gauge lengths, so it does not provide any improvement over this latter model (see Figure 1 in the  Supplementary Material).
Note that, since the sample sizes are around 150, the value of \(\epsilon^*(0.05) \) (resp. \(G^*(F,\mathcal{G})\)) obtained with the Bootstrap 1 method is noticeably lower (resp. higher) than the one derived with the Bootstrap 2: this was also the case in the simulations.
We have carried out a linear regression of \(G^*(F,\mathcal{G})\) over the gauge length for each model, metric and bootstrap procedure (a summary of the results appears in Table~\ref{Tab.FailureStress4}) (see also Figure 2 in the Supplementary Material). Observe that the percentage of improvement of the Weibull, the 3-Weibull and the skew normal models over a constant one for the failure stress decreases significantly (at the 5\% significance level in all cases) as the gauge length increases. In contrast, at the 5\% level, that percentage of improvement is not linearly dependent of the gauge length for the bimodal Weibull. As a matter of fact, this last model attains the highest value of \(G^*(F,\mathcal{G})\) for the majority (6 or 7 over 8) of gauge lengths in all cases. As a conclusion, we consider that, among the considered models, the mixture of two Weibulls is the distribution providing the best fit to this failure stress sample.

\begin{table}[h!]
\begin{center}
\begin{tabular}{c@{\hspace{2mm}}c@{\hspace{3mm}}c@{\hspace{3mm}}cccc}
Gauge & & & \multicolumn{2}{c}{\(\mathrm{L}^1\)-distance} & \multicolumn{2}{c}{\(\mathrm{L}^2\)-distance} \\
length & \(n\) & Model & Bootstrap 1 & Bootstrap 2 & Bootstrap 1 & Bootstrap 2 \\ \hline
20 & 153 & W & 47.61 (0.926) & 74.49 (0.884) & 0.8621 (0.937) & 1.3892 (0.899) \\
   &    & 3W & 48.36 (0.925) & 74.71 (0.884) & 0.8704 (0.937) & 1.3917 (0.899) \\
   &    & SN & 36.51 (0.944) & 60.15 (0.907) & 0.7116 (0.948) & 1.1908 (0.914) \\
   &    & BW & 23.36 (0.964) & 86.14 (0.867) & 0.4743 (0.966) & 1.7272 (0.875) \\ \hline
30 & 151 & W & 47.30 (0.915) & 65.67 (0.882) & 1.0677 (0.918) & 1.4664 (0.888) \\
   &    & 3W & 47.07 (0.915) & 62.63 (0.887) & 1.0104 (0.923) & 1.3651 (0.896) \\
   &    & SN & 53.04 (0.905) & 70.87 (0.873) & 1.1406 (0.913) & 1.5721 (0.880) \\
   &    & BW & 43.28 (0.922) & 52.48 (0.906) & 0.8700 (0.934) & 1.1136 (0.915) \\ \hline
40 & 149 & W & 40.00 (0.925) & 59.89 (0.888) & 0.9109 (0.929) & 1.3529 (0.895) \\
   &    & 3W & 41.02 (0.923) & 60.17 (0.887) & 0.9278 (0.928) & 1.3692 (0.894) \\
   &    & SN & 40.14 (0.925) & 56.47 (0.894) & 0.8550 (0.934) & 1.2379 (0.904) \\
   &    & BW & 48.29 (0.910) & 53.66 (0.900) & 1.0508 (0.919) & 1.1904 (0.908) \\ \hline
45 & 153 & W & 37.20 (0.932) & 60.71 (0.890) & 0.8497 (0.934) & 1.3292 (0.897) \\
   &    & 3W & 38.21 (0.931) & 61.07 (0.889) & 0.8554 (0.934) & 1.3330 (0.897) \\
   &    & SN & 31.97 (0.942) & 51.73 (0.906) & 0.6456 (0.950) & 1.0999 (0.915) \\
   &    & BW & 31.26 (0.943) & 43.48 (0.921) & 0.6088 (0.953) & 0.9037 (0.930) \\ \hline
50 & 152 & W & 56.73 (0.905) & 80.52 (0.866) & 1.1398 (0.912) & 1.6036 (0.876) \\
   &    & 3W & 57.07 (0.905) & 80.57 (0.865) & 1.1395 (0.912) & 1.6012 (0.877) \\
   &    & SN & 47.62 (0.920) & 69.84 (0.883) & 0.9525 (0.927) & 1.3947 (0.893) \\
   &    & BW & 44.78 (0.925) & 59.28 (0.901) & 0.9170 (0.929) & 1.2189 (0.906) \\ \hline
60 & 151 & W & 60.76 (0.875) & 81.71 (0.833) & 1.1144 (0.905) & 1.5614 (0.867) \\
   &    & 3W & 66.47 (0.864) & 84.71 (0.827) & 1.1708 (0.900) & 1.6094 (0.863) \\
   &    & SN & 60.19 (0.877) & 80.61 (0.835) & 1.0817 (0.908) & 1.5155 (0.871) \\
   &    & BW & 61.95 (0.873) & 66.97 (0.863) & 1.2839 (0.890) & 1.4019 (0.881) \\ \hline
65 & 151 & W & 50.08 (0.890) & 67.03 (0.854) & 1.1386 (0.899) & 1.5370 (0.863) \\
   &    & 3W & 51.93 (0.887) & 65.78 (0.857) & 1.1514 (0.898) & 1.4942 (0.867) \\
   &    & SN & 43.64 (0.905) & 59.21 (0.871) & 0.9799 (0.913) & 1.3393 (0.881) \\
   &    & BW & 19.40 (0.958) & 40.07 (0.913) & 0.3543 (0.968) & 0.8703 (0.923) \\ \hline
80 & 153 & W & 68.34 (0.871) & 96.81 (0.818) & 1.1964 (0.903) & 1.7237 (0.860) \\
   &    & 3W & 68.84 (0.871) & 97.09 (0.818) & 1.1985 (0.903) & 1.7265 (0.860) \\
   &    & SN & 79.09 (0.852) & 94.79 (0.822) & 1.3813 (0.888) & 1.7743 (0.856) \\
   &    & BW & 35.82 (0.933) & 53.57 (0.899) & 0.9503 (0.923) & 1.2546 (0.898) \\ \hline
\end{tabular}
\end{center}
\caption{For the failure stress data and for each gauge length (column 1), sample size (column 2), parametric model (column 3), \(\epsilon_k^*(0.05)\) (columns 4--7) and, between parentheses, \({G}^*(F,\mathcal{G}_k)\).}
\label{Tab.FailureStress3}
\end{table}

\begin{table}[h!]
\begin{center}
\begin{tabular}{cccccc}
Distance & Procedure & Model & Slope & p-value & Correlation \\ \hline
\(\mathrm{L}^1\) & Bootstrap 1 &  W & $-0.0010$ & 0.0096 & $-0.84$ \\
                 &             & 3W & $-0.0011$ & 0.0134 & $-0.82$ \\
                 &             & SN & $-0.0013$ & 0.0208 & $-0.79$ \\
                 &             & BW & $-0.0003$ & 0.6230 & $-0.21$ \\ \hline
                 & Bootstrap 2 &  W & $-0.0012$ & 0.0059 & $-0.86$ \\
                 &             & 3W & $-0.0013$ & 0.0074 & $-0.85$ \\
                 &             & SN & $-0.0013$ & 0.0222 & $-0.78$ \\
                 &             & BW &  0.0002 & 0.6460 &  0.19 \\ \hline
\(\mathrm{L}^2\) & Bootstrap 1 &  W & $-0.0006$ & 0.0130 & $-0.82$ \\
                 &             & 3W & $-0.0007$ & 0.0085 & $-0.84$ \\
                 &             & SN & $-0.0008$ & 0.0327 & $-0.75$ \\
                 &             & BW & $-0.0004$ & 0.4500 & $-0.31$ \\ \hline
                 & Bootstrap 2 &  W & $-0.0007$ & 0.0031 & $-0.89$ \\
                 &             & 3W & $-0.0008$ & 0.0016 & $-0.91$ \\
                 &             & SN & $-0.0008$ & 0.0331 & $-0.75$ \\
                 &             & BW &  0.0001 & 0.7410 &  0.14 \\ \hline
\end{tabular}
\end{center}
\caption{For the failure stress data, the linear regression parameter (column 2) of \({G}^*(F,\mathcal{G})\) over the gauge length, the p-value of the regression test  (column 3) and the correlation between \({G}^*(F,\mathcal{G})\) and the gauge length.}
\label{Tab.FailureStress4}
\end{table}


\section{Discussion}

The objective of this paper is to determine whether a parametric model provides a sufficiently good fit to the observed data. For this purpose, we introduce the AGoF test by which we can decide whether the unknown distribution of the data is within a certain margin of error of the proposed model in terms of the \(\mathrm{L}^p\)-distance. The value of $p$ can be chosen a priori by the data analyst depending on the importance of the tails of the distribution in the problem under consideration. Our strategy differs from others considered in the literature. In the alternative hypothesis we handle full topological neighborhoods of a suitable representative from the parametric class in the model. Other approaches only consider two fixed cdf (\cite{Munk-Czado-1998}), smaller alternatives (\cite{Liu-Lindsay-2009}) or contamination neighborhoods (\cite{AE-B-CA-M-2012} and \cite{del Barrio-2020}), without considering parametric families, which seems to be of more practical relevance.

The choice of a specific value for the margin of error $\epsilon$, a delicate issue in this type of tests, is avoided by determining the smallest distance at which $H_0$ is rejected at significance level $\alpha$. Another contribution of this work is the introduction of the AGoF statistic in \eqref{Proportion-improvement} to quantify the proportion of the observed variable explained by the model in comparison to a constant, non-informative, one. In this way, different parametric models can be easily compared by simply ordering the values of this quantity. We can also assess whether a more complex model provides sufficient improvement over a simpler one and interpret values of distances between distributions with metrics that are not as intuitive as the usual supremum norm.

To carry out the test, we propose two possible methods based on bootstrap and supported by the developed asymptotic theory. We determine the conditions under which the bootstrap estimators are consistent and check the performance of the methodology by means of simulations. Based on the results from this Monte Carlo study and our theoretical results, we give a recommendation for the use of each of these two computational methods.

For future work, it would be interesting to extend the test to the multivariate context. This extension would require the use of a suitable and easy-to-handle (functional) metric between probability distributions.



\section*{Acknowledgements}

We thank the Centers for Disease Control and Prevention (United States) and the Minist\`{e}re de la Sant\'{e} Publique et de la Population (Haiti) for providing the IgG antibody data from the Haiti serosurvey (Section~\ref{Section.AntibodiesHaiti}).
We would also like to express our gratitude to the reviewers for their careful reading of the first version of the manuscript, and for pointing out the references \cite{Baringhaus-Henze}, \cite{Berger-Delampady}, and \cite{Dette-Sen}, which have been incorporated into the revised version.
A. Ba\'illo and J. C\'{a}rcamo are supported by the Spanish MCyT grant PID2023-148081NB-I00.



\newpage
\section*{Appendix (Proofs of the mathematical results)}

We first observe that condition (\ref{asymptotically-linear}) means that $\{\hat{\boldsymbol{\theta}}_n\}$ is asymptotically linear at $\boldsymbol{\theta}_F$. The differentiability of the map  $\boldsymbol{\theta}\mapsto \E_F {\boldsymbol{\psi}}_{\boldsymbol{\theta}}(X)$ and the fact that $\boldsymbol{V}_{\boldsymbol{\theta}_F}$ is invertible allows for expanding ${\boldsymbol{\Psi}}_n$ in (\ref{Psi}) around ${\boldsymbol{\theta}}_F$ and solving the factor $(\hat{\boldsymbol{\theta}}_n-\boldsymbol{\theta}_F)$. Finally, the requirement $\E_F\Vert \boldsymbol{\psi}_{\boldsymbol{\theta}_F}(X)\Vert^2<\infty$ is necessary to apply the usual CLT to the sum in (\ref{asymptotically-linear}). We observe that, as ${\boldsymbol{\theta}}_F$ is in the interior of $\Theta$, condition (\ref{asymptotically-linear}) implies that $\P(\hat{\boldsymbol{\theta}}_n \in \Theta)\to 1$, as $n\to \infty$. Therefore, we can always assume that $\hat{\boldsymbol{\theta}}_n \in \Theta$. 

We consider
\begin{equation} \label{empirical process}
\Emp_n(t) = \sqrt{n}(\mathbb{F}_n(t)-F(t)),\qquad n\in \mathbb{N},\quad t\in \R,
\end{equation}
the \emph{empirical process} associated to the sample \(X_1,\ldots,X_n\) from \(F\). Let $ \mathbb{X}_1,\dots, \mathbb{X}_n$ be independent copies of the process
\begin{equation}\label{Empirical X}
\mathbb{X}(t)=\P(X>t)-1_{\{X>t\}},\qquad t\in \R.
\end{equation}
Then, the empirical process \eqref{empirical process} can be expressed as
\begin{equation} \label{Empirical representation} 
\Emp_n(t) 
= \frac{1}{\sqrt{n}}\sum_{i=1}^n \mathbb{X}_i,\quad n\in \mathbb{N},\quad t\in \R.
\end{equation}
To prove Theorem~\ref{Theorem.Process.parameters} we need to establish in advance the conditions under which the empirical process converges weakly to the \(F\)-Brownian bridge in $\mathrm{L}^p$.
The results
in \cite{Araujo-Gine} and \cite{Ledoux-Talagrand} regarding when a random variable (such as \(\mathbb{X}\)) taking values in a Banach space satisfies the Central Limit Theorem (CLT) distinguish between cotype 2 and type 2 spaces. For $1\le p\le 2$, $\mathrm{L}^p$ has cotype 2 and, for $2<p<\infty$, $\mathrm{L}^p$ has type 2 and satisfies the Rosenthal property (see \cite{Ledoux-Talagrand}). As a consequence, we have the following characterizations:
\begin{itemize}
\item[--] If $1\le p\le 2$, a centered r.v. \(\mathbb{X}\) with values in $\mathrm{L}^p$ satisfies the CLT if and only if it is pregaussian (see, e.g., \cite{Ledoux-Talagrand} for a definition of pregaussian).
\item[--] If $2<p<\infty$, $\mathbb{X}$ satisfies the CLT in $\mathrm{L}^p$ if and only if $\mathbb{X}$ is pregaussian and satisfies
\begin{equation}\label{tail condition}
\lim_{t\to\infty}t^2\P(\Vert\mathbb{X} \Vert_{p}>t)=0.
\end{equation}
\end{itemize}
Further, \cite{Ledoux-Talagrand} state that a centered $\mathrm{L}^p$-valued random variable $\mathbb{X}$ is pregaussian if and only if
\begin{equation}\label{CLT pregaussian}
\int_{\R} (\E \mathbb{X}^2(t))^{p/2} \, \d t <\infty.
\end{equation}

Theorem A.\ref{Theorem main} gives a characterization of when \(\mathbb{X}\) satisfies the CLT in \(\mathrm{L}^p\). It is used as an auxiliary result to prove Theorem~\ref{Theorem.Process.parameters}. For the proof of Theorem A.\ref{Theorem main} and the rest of this section we use the notation $S_F=\{ t\in \R : F(t)\in(0,1)\}.$

\begin{theoremAppendix}\label{Theorem main}
For $1\le p <\infty$, the following assertions are equivalent.
\begin{enumerate}[label=(\alph*), topsep=0mm, itemsep=1mm, parsep=0mm, align=left, labelwidth=*, leftmargin=-0.5 mm]
\item $\Emp_n\cw \B_F$ in $\mathrm{L}^p$.
\item $\Emp_n$ is pregaussian in $\mathrm{L}^p$.
\end{enumerate}
\end{theoremAppendix}

\begin{proof} When $1\le p\le 2$, the result is precisely the characterization given above for cotype 2 spaces. Thus, from now on, we assume that $2<p<\infty$.
Also by the above characterizations, we know that condition (a) always implies condition (b). Then it only remains to prove that (b) implies (a). From (\ref{Empirical representation}), we see that (b) is equivalent to the process $\mathbb{X}$ in (\ref{Empirical X}) being pregaussian. In turn, this is equivalent to $\mathbb{X}$ fulfilling \eqref{CLT pregaussian}.
It is straighforward to see that \eqref{CLT pregaussian} is equivalent to
\begin{equation}\label{Psi_p}
\int_{\R} \, \left[ F(t)\bar{F}(t) \right]^{p/2}\d t <\infty,
\end{equation}
where $\bar{F}\equiv 1-F$ denotes the survival function of $X$.
By \eqref{Psi_p}, the paths of \(\Emp_n\) are in $\mathrm{L}^p$ a.s. Observe also that, when the empirical process converges, its only possible Gaussian limit is \(\B_F\) (as their covariances coincide).

In addition, condition \eqref{Psi_p} is equivalent to stating that, for any $c\in S_F$, we have
\begin{equation*}
\int_0^\infty \P \{|X-c|>t\}^{p/2} \, \d t<\infty,
\end{equation*}
which implies that
\begin{equation} \label{Lemma3Manuscript}
\P \{|X-c|>t\} = o(t^{-2/p}),\quad \mbox{as $t\to\infty$}.
\end{equation}
It only remains to check that \eqref{Lemma3Manuscript} implies \eqref{tail condition}. Observe that, for any $c\in S_F$,
\begin{eqnarray}
\Vert \mathbb{X} \Vert_p^p & = & \int_{(-\infty,X)} F(t)^p\,\d t+\int_{[X,\infty)} \bar F(t)^p\, \d t \nonumber \\
& \geq & m_c^p \, |X-c| , \label{norm of X}
\end{eqnarray}
where \( m_c = \min\{F(c),\bar F(c)\}\). Consequently,
\[
\P \{ |X-c|>t^p\} \leq \P \{\Vert \mathbb{X} \Vert_p > m_c \, t \}
\]
and this last inequality, together with \eqref{Lemma3Manuscript}, yields \eqref{tail condition}.
\end{proof}

Now we can prove Theorem~\ref{Theorem.Process.parameters}. To this end, we say that two processes, $\mathbb{P}_n$ and $\widetilde{\mathbb{P}}_n$, taking values in $\mathrm{L}^p$ a.s., are \textit{equivalent} in $\mathrm{L}^p$ if $\Vert \mathbb{P}_n-\widetilde{\mathbb{P}}_n\Vert_p\cp 0$. Note that if  $\mathbb{P}_n$ and $\widetilde{\mathbb{P}}_n$ are equivalent in $\mathrm{L}^p$ and $\mathbb{P}_n\cw \mathbb{P}$ in $\mathrm{L}^p$ then $\widetilde{\mathbb{P}}_n\cw \mathbb{P}$ in $\mathrm{L}^p$ (see \cite{van der Vaart}).

{\em Proof of Theorem~\ref{Theorem.Process.parameters}.} Assumptions 1 and 2 imply that $\mathbb{G}_n({\boldtheta}_F)$ in \eqref{Process-Gn(theta)} is equivalent in $\mathrm{L}^p$ to
\begin{equation*}
\mathbb{G}_n^*({\boldtheta}_F)=\sqrt{n}(\mathbb F_n- F) - \sqrt{n}(\hat{\boldsymbol{\theta}}_n-\boldsymbol{\theta}_F)^T  \dot{\G}(\boldsymbol{\theta}_F).
\end{equation*}

Let us check next that $\mathbb{G}_n^*$ is equivalent in $\mathrm{L}^p$ to the process
\begin{equation}\label{equivalent-process}
\tilde{\mathbb{G}}_n ({\boldtheta}_F)=\sqrt{n}(\bbF_n- F) - \frac{1}{\sqrt{n}}\sum_{i=1}^n  \mathbf{l}_{{\boldsymbol{\theta}_F}}(X_i)^T \dot{\G}(\boldsymbol{\theta}_F).
\end{equation}
Denoting
\begin{equation*}
\Vert \dot{\G}(\boldsymbol{\theta}_F) \Vert_p
= (  \Vert \dot{G}_1( {\boldtheta}_F )\Vert_p,\dots,\Vert \dot{G}_k({\boldtheta}_F)\Vert_p )^T \quad \text{and}\quad |\mathbf{v}|
= (|v_1|,\dots,|v_k|)^T,
\end{equation*}
for a vector $\mathbf{v}\in \R^k$, by Minkowski inequality, we obtain that
\begin{equation*}
\Vert \mathbb{G}_n^*({\boldtheta}_F) - \tilde{\mathbb{G}}_n ({\boldtheta}_F) \Vert_p
\le \left| \sqrt{n}(\hat{\boldsymbol{\theta}}_n-\boldsymbol{\theta}_F)
- \frac{1}{\sqrt{n}}\sum_{i=1}^n \mathbf{l}_{{\boldsymbol{\theta}_F}}(X_i) \right|^T  \Vert \dot{\G}(\boldsymbol{\theta}_F) \Vert_p ,
\end{equation*}
and this last quantity is $o_\P(1)$ by \eqref{asymptotically-linear} and by Assumption 1. We conclude that $\mathbb{G}_n({\boldtheta}_F)$ in \eqref{Process-Gn(theta)} and $\tilde{\mathbb{G}}_n ({\boldtheta}_F)$ in \eqref{equivalent-process} are equivalent in $\mathrm{L}^p$.

Now, $\tilde{\mathbb{G}}_n ({\boldtheta}_F)$ can be written in a normalized form as
\begin{equation*}
\tilde{\mathbb{G}}_n ({\boldtheta}_F) = \frac{1}{\sqrt{n}} \sum_{i=1}^n \mathbb{Z}_i,
\end{equation*}
where $ \mathbb{Z}_1,\dots, \mathbb{Z}_n$ are independent copies of the process
\begin{equation}\label{Empirical estimated parameters Z}
\mathbb{Z}= \mathbb{X} - \mathbf{l}_{{\boldsymbol{\theta}_F}}(X)^T  \dot{\G}(\boldsymbol{\theta}_F).
\end{equation}
Therefore, to finish the proof of the theorem we have to prove that $\mathbb{X}$ satisfies the CLT in $\mathrm{L}^p$ if and only if  $\mathbb{Z}$ satisfies the CLT in $\mathrm{L}^p$. 

Let us first assume that Theorem \ref{Theorem.Process.parameters} (i) holds, i.e., $\mathbb{X}$ satisfies the CLT in $\mathrm{L}^p$. By Minkowski inequality, we obtain that
\begin{equation} \label{IneqLpNormZ}
\Vert \mathbb{Z} \Vert_p \le \Vert \mathbb{X} \Vert_p + \left|\mathbf{l}_{{\boldsymbol{\theta}_F}}(X)\right|^T \Vert \dot{\G}(\boldsymbol{\theta}_F) \Vert_p.
\end{equation}
By Assumptions 1 and 2, the random variable $Y = \left| \mathbf{l}_{{\boldsymbol{\theta}_F}}(X) \right|^T \Vert \dot{\G}(\boldsymbol{\theta}_F) \Vert_p$ has finite second moment, and hence, $\P( Y > t)=o(t^{-2})$, as $t\to\infty$. Therefore, we conclude that $\P( \Vert \mathbb{Z}  \Vert_p > t)=o(t^{-2})$, as $t\to\infty$, if \(2<p<\infty\). Additionally, by Cauchy–Schwarz inequality,
\begin{equation*}
\mathbb{Z}^2 \le 2\left( \mathbb{X}^2 +\Vert \mathbf{l}_{{\boldsymbol{\theta}_F}}(X) \Vert^2   \cdot  \Vert \dot{\G}(\boldsymbol{\theta}_F) \Vert^2  \right),
\end{equation*}
where we recall that $\Vert \cdot\Vert $ is the Euclidean norm in $\mathbb{R}^k$.
Therefore,
\begin{equation}\label{bound Z X}
(\E {\mathbb{Z}}^2)^{p/2} \le 2^p\left(    (\E {\mathbb{X}}^2)^{p/2}   +   \left(\E \Vert \boldsymbol{V}_{\boldsymbol{\theta}_F}^{-1} \cdot \boldsymbol{\psi}_{\boldsymbol{\theta}_F}(X)  \Vert^2\right)^{p/2} \cdot  \Vert \dot{\G}(\boldsymbol{\theta}_F) \Vert^p  \right).
\end{equation}
Using \eqref{CLT pregaussian} and
\begin{equation*}
\Vert \dot{\G}(\boldsymbol{\theta}_F) \Vert^p\le k^{p/2} \left(    |\dot{G}_1( {\boldtheta}_F )|^p +\cdots +  |\dot{G}_d( {\boldtheta}_F )|^p  \right)\in \mathrm{L}^1,
\end{equation*}
by (\ref{bound Z X}) we have that $\int \left(\E\mathbb{Z}(t)^2\right)^{p/2} \, \d t<\infty$ and part (ii) holds.

Conversely, assume that Theorem \ref{Theorem.Process.parameters} (ii) is satisfied. In other words, the variable $\mathbb{Z}$ in \eqref{Empirical estimated parameters Z} satisfies the CLT in $\mathrm{L}^p$. In particular, $\mathbb{Z}$ is pregaussian in $\mathrm{L}^p$, that is,  $\int \left(\E\mathbb{Z}(t)^2\right)^{p/2} \, \d t<\infty$. Now, by \eqref{Empirical estimated parameters Z}, $\mathbb{X}=\mathbb{Z}+\mathbf{l}_{{\boldsymbol{\theta}_F}}(X)^T  \dot{\G}(\boldsymbol{\theta}_F)$.
Following the same lines as above, we obtain that
\begin{equation*}
(\E {\mathbb{X}}^2)^{p/2} \le 2^p\left(  (\E {\mathbb{Z}}^2)^{p/2}  +  \left(\E \Vert \boldsymbol{V}_{\boldsymbol{\theta}_F}^{-1} \cdot \boldsymbol{\psi}_{\boldsymbol{\theta}_F}(X)  \Vert^2\right)^{p/2} \cdot  \Vert \dot{\G}(\boldsymbol{\theta}_F) \Vert^p  \right) \in \mathrm{L}^1.
\end{equation*}
Therefore, $\mathbb{X}$ is pregaussian in $\mathrm{L}^p$ and, by Theorem A.\ref{Theorem main}, $\mathbb{X}$ satisfies the CLT in $\mathrm{L}^p$ and the proof of the theorem is complete. \hfill $\square$
\medskip

{\em Proof of Theorem~\ref{Theorem-probability-metric-parameters}.}
By the proof of Theorem A.\ref{Theorem main}, the assumption that $X\in\mathcal{L}^{2/p,1}$ is equivalent to $\Emp_n\cw \B_F$ in $\mathrm{L}^p$, which, in turn is equivalent to $\mathbb{G}_n({\boldtheta}_F)\cw \mathbb{G}_{{\boldtheta}_F}$ in  $\mathrm{L}^p$ (by Theorem~\ref{Theorem.Process.parameters}). Now, the normalized test statistic in \eqref{normalized-statistic} can be written as
\begin{equation}\label{Tn-Delta-method}
    T_{n}(F, G({\boldtheta}_F), p) = \sqrt{n} \left( \delta_p(\bbF_n - G(\hat\btheta_n)) - \delta_p(F - G(\btheta_F)) \right),
\end{equation}
where $\delta_p(f) = \|f\|_p$ denotes the $\mathrm{L}^p$-norm of a function $f \in \mathrm{L}^p(\mathbb{R})$.

Note that $T_{n}(F, G({\boldtheta}_F), p)$ is now expressed in a form suitable for applying the functional delta method. First, the map $\delta_p(\cdot)$ is directionally Hadamard differentiable, as shown in \citet[Lemma~4]{Carcamo}. Therefore, we may apply the extended version of the functional delta method (see \cite{Shapiro} or \cite{Fang-Santos}) to obtain
\[
T_{n}(F, G({\boldtheta}_F), p) \cw (\delta_p)'_{F - G(\btheta_F)} (\mathbb{G}_{{\boldtheta}_F}),
\]
where the expression for the directional derivative $(\delta_p)'_{F - G(\btheta_F)}$ is given in \citet[Lemma~4]{Carcamo}. This completes the proof of the theorem. \hfill $\Box$

\medskip

{\em Proof of Corollary~\ref{Corollary normalidad}.}
Since \(\mathbb{G}_{{\boldtheta}_F}\) is a centered Gaussian process and \(F-G({\boldtheta}_F)\) is non-random, the integrals
\[
\int_{\R\setminus {C_{{\boldtheta}_F}}} \mathbb{G}_{{\boldtheta}_F} \, \sgn(F-G({\boldtheta}_F)) \quad \mbox{and} \quad \int \mathbb{G}_{{\boldtheta}_F} \, |F-G({\boldtheta}_F)|^{p-1}\,\sgn(F-G({\boldtheta}_F)),
\]
appearing, respectively, in the representation of \(T(F,G({\boldtheta}_F),1)\) in \eqref{AsympDistr_p_equal_1} and of \(T(F,G({\boldtheta}_F),p)\), for \(1<p<\infty\), in \eqref{AsympDistr_p_larger_1}, have zero-mean Gaussian distribution. Thus, \(T(F,G({\boldtheta}_F),1)\) is a zero-mean normal if and only if \(C_{{\boldtheta}_F}\) has zero Lebesgue measure. The case \(1<p<\infty\) follows from Theorem~\ref{Theorem-probability-metric-parameters}(b). \hfill $\Box$
\medskip

{\em Proof of Theorem~\ref{Theorem.Bootstrap.Consistency}.}
By Assumptions 1, 3 and 4, the process \(\bbG_n^*(\hat\btheta_n)\) in \eqref{EmpProcEstParBOOT} is equivalent in \(\mathrm{L}^p\) to the process
\[
\sqrt{n} (\bbF_n^*-\bbF_n) - \sqrt{n} \, \dot{\mathbf{G}}(\btheta_F)^T (\hat\btheta_n^*-\hat\btheta_n),
\]
which in turn, by Assumption 5, is equivalent in \(\mathrm{L}^p\) to
\begin{equation} \label{EmpProcEstParBOOTtilde}
\tilde\bbG_n^*  =  
\sqrt{n} (\bbF_n^*-\bbF_n)
 + \dot{\mathbf{G}}(\btheta_F)^T \bV_{\btheta_F}^{-1} \sqrt{n} (\bbF_n^*-\bbF_n)(\bpsi_{\btheta_F}).
\end{equation}
Consequently, it suffices to prove that
\begin{equation} \label{TildeGStar}
\tilde\bbG_n^* \cw \bbG_{\btheta_F}  \quad \mbox{in } \mathrm{L}^p \quad F\mbox{-a.s.}
\end{equation}

Observe that
\begin{equation*}
\tilde\bbG_n^*(t) 
  =  \frac{1}{\sqrt{n}} \sum_{i=1}^n (M_{ni}-1) \bbZ_i(t),
\end{equation*}
where \(M_{ni}\) denotes the absolute frequency of \(X_i\) in the bootstrap sample and $\mathbb{Z}_1,\dots,\mathbb{Z}_n$ are independent copies of the process
\eqref{Empirical estimated parameters Z}.

Since \(\sum_{i=1}^n M_{ni}=n\), the multipliers \(M_{ni}-1\) are dependent. First, we remove this dependence by Poissonization (see \cite[Section 3.7.1]{van der Vaart-Wellner 2023}) as follows.
Instead of \(n\), let the bootstrap sample size be  \(N_n\), a Poisson r.v. independent of \(X_1,\ldots,X_n\) and with mean \(n\). The absolute frequency of \(X_i\) in the bootstrap sample with size \(N_n\) is replaced by \(M_{N_n,i}\), where, for \(i=1,\ldots,n\), \(M_{N_n,i}\) are independent Poisson r.v. with mean 1.
Define
\[
\tilde\bbZ_n(t) = 
\frac{1}{\sqrt{n}} \sum_{i=1}^n (M_{N_n,i}-1) \bbZ_i(t).
\]
By \cite[Lemma 1.10.2 (i)]{van der Vaart-Wellner 2023}, to prove \eqref{TildeGStar} it suffices to see that, with \(F\)-probability 1 (i.e., for almost all sequences \(X_1,X_2,\ldots\)), the following points (i) and (ii) are satisfied:
\begin{enumerate}[label=(\roman*), topsep=0mm, itemsep=1mm, parsep=0mm, align=left, labelwidth=*, leftmargin=-0.5 mm]
\item the process \(\tilde\bbZ_n\) converges weakly to \(\bbG_{\btheta_F}\) in \(\mathrm{L}^p\);
\item \(\tilde\bbG_n^*\) and \(\tilde\bbZ_n\) are equivalent in $\mathrm{L}^p$.
\end{enumerate}

(i)
By \cite[Thm. 10.14]{Ledoux-Talagrand},
the weak convergence of the process \(\tilde\bbG_n(\btheta_F)\) in \eqref{equivalent-process}
to \(\bbG_{\btheta_F}\) in \(\mathrm{L}^p\), together with \(\E \|\bbZ\|_p^2 <\infty\), is equivalent to the weak convergence of \(\tilde\bbZ_n\) to \(\bbG_{\btheta_F}\) in \(\mathrm{L}^p\), for almost every sequence \(X_1,X_2,\ldots\)
By inequality \eqref{IneqLpNormZ} and the fact that the variable $Y = \left| \mathbf{l}_{{\boldsymbol{\theta}_F}}(X) \right|^T \Vert \dot{\G}(\boldsymbol{\theta}_F) \Vert_p$ has finite second moment (see the proof of Theorem~\ref{Theorem.Process.parameters}), to check that \(\E \|\bbZ\|_p^2 <\infty\) it suffices to prove that
\(\E \|\bbX\|_p^2 <\infty\). Now, it can be seen that \(\|\bbX\|_p \leq |X|^{1/p} + \|\tilde F\|_p\), where \(\tilde F(t)=F(t)\) if \(t<0\) and \(\tilde F(t)=\bar F(t)\) if \(t\geq0\). Since the integrability conditions on $X$ imply that \(X\in\mathcal{L}^{2/p}\), we conclude that \(\E \|\bbX\|_p^2 <\infty\).

(ii)
We have to check that, for all \(\epsilon>0\),
\begin{equation} \label{DistPoissOrigBoot}
\P \left\{ \|\tilde\bbZ_n - \tilde\bbG_n^*\|_p >\epsilon \mid X_1,\dots,X_n \right\} \to 0 \quad F\mbox{-a.s.},
\end{equation}
where the probability is taken with respect to the resampling mechanism and the Poisson r.v.'s \(N_n\) and \(M_{N_n,i}\), \(i=1,\ldots,n\).
First observe that
\[
\tilde\bbZ_n - \tilde\bbG_n^* = \frac{1}{\sqrt{n}} \sum_{i=1}^n (M_{N_n,i}-M_{ni}) \bbZ_i.
\]
Denote by \(I_n^j\) the set of indices \(i\in\{1,2,\ldots,n\}\) such that \(|M_{N_n,i}-M_{ni}|\geq j\). Then
\[
M_{N_n,i}-M_{ni} = \mbox{sgn}(N_n-n) \sum_{j=1}^\infty 1_{\{i\in I_n^j\}}.
\]
We have that
\begin{align*}
\tilde\bbZ_n - \tilde\bbG_n^* & =
\mbox{sgn}(N_n-n) \sum_{j=1}^\infty \frac{1}{\sqrt{n}} \sum_{i=1}^n 1_{\{i\in I_n^j\}} \bbZ_i \\
& =  
\mbox{sgn}(N_n-n) \sum_{j=1}^\infty \frac{\# I_n^j}{\sqrt{n}} \left( \frac{1}{\# I_n^j} \sum_{i\in I_n^j} \bbZ_i \right).
\end{align*}
Now, we consider the event \( B= \{ \max_{1\leq i\leq n}|M_{N_n,i}-M_{ni}|>2 \} \). We have that
\[
\P \left\{ \|\tilde\bbZ_n - \tilde\bbG_n^*\|_p >\epsilon \right\} \leq \P (B)
+\P (B^c) \, \P \left\{ \|\tilde\bbZ_n - \tilde\bbG_n^*\|_p >\epsilon \; \left| B^c \right. \right\} .
\]
In \cite[pp.~494--495]{van der Vaart-Wellner 2023} it is proved that, for every \(\delta>0\),
\( \P (B) \to \delta \), as \(n\to\infty\).
This entails that, for sufficiently large \(n\), with probability at least \(1-2\delta\), all the terms \(|M_{N_n,i}-M_{ni}|\) are 0, 1 or 2. Consequently, it remains to prove that, for \(j=1\) and 2 and for all \(\epsilon>0\), it holds that
\begin{equation} \label{ConvProbAS}
\P \left\{ \frac{\# I_n^j}{\sqrt{n}} \left\| \frac{1}{\# I_n^j} \sum_{i\in I_n^j} \bbZ_i \right\|_p  >\epsilon \right\} \to 0 \quad F\mbox{-a.s.}
\end{equation}
In \cite[pp.~494--495]{van der Vaart-Wellner 2023} it is noted that \(j(\# I_n^j) \leq |N_n-n| = O_{\P}(\sqrt{n})\). So \eqref{ConvProbAS} reduces to proving
\[
\P \left\{ \left\| \frac{1}{\# I_n^j} \sum_{i\in I_n^j} \bbZ_i \right\|_p >\epsilon \right\} \to 0 \quad F\mbox{-a.s.}
\]
This convergence is obtained by applying Lemma A.\ref{Lemma_3_6_16_vdV_W} below with weights \(W_{ni}=1_{\{i\in I_n^j\}}/\# I_n^j\).
Its proof is analogous to that in \citet[Lemma 3.7.16]{van der Vaart-Wellner 2023}, substituting the sup-norm by the \(\mathrm{L}^p\)-norm.
\hfill \(\Box\)

\begin{lemmaAppendix} \label{Lemma_3_6_16_vdV_W}
For each \(n\), let \((W_{n1},\ldots,W_{nn})\) be exchangeable non-negative r.v. independent of \(X_1,\ldots,X_n\) such that \(\sum_{i=1}^n W_{ni}=1\) and \(\max_{1\leq i\leq n} W_{ni} \cp 0\). Assume that \(X\in\mathcal L^{2/p}\). Then, under Assumptions 1 and 2, for every \(\epsilon>0\), as \(n\to\infty\), we have that
\[
\P_W \left\{ \left\| \sum_{i=1}^n W_{ni} \bbZ_i \right\|_p>\epsilon \right\} \to 0 \qquad F\mbox{-a.s.}
\]
\end{lemmaAppendix}


\end{document}